\documentclass[a4paper,11pt]{JHEP}
\usepackage{graphicx}
\usepackage{amsmath}
\usepackage{amssymb}
\usepackage{multirow}

\newcommand{\url}[1]{\href{#1}{#1}}

\preprint{CERN-PH-TH/2012-261, LAPTH-046/12}

\title{Neutrino masses and cosmological parameters from a Euclid-like survey: Markov Chain Monte Carlo forecasts including
  theoretical errors}

\author{
Benjamin Audren$^a$, 
Julien Lesgourgues$^{a,b,c}$,
Simeon Bird$^d$,
Martin G. Haehnelt$^e$,
Matteo Viel$^{f,g}$\\ 
{$^a$}Institut de Th\'eorie des Ph\'enom\`enes Physiques,\\ 
\'Ecole PolytechniqueF\'ed\'erale de Lausanne, 
CH-1015, Lausanne,Switzerland.\vspace{.2cm}\\ 
{$^b$} CERN, Theory Division,\\ 
CH-1211 Geneva 23, Switzerland.\vspace{.2cm}\\ 
{$^c$} LAPTh (CNRS - Universit\'e de Savoie), BP 110,\\ F-74941 Annecy-le-Vieux Cedex, France.\vspace{.2cm}\\
{$^d$}Institute for Advanced Study,\\ 
1 Einstein Drive, 
Princeton, NJ, 08540, USA.\vspace{.2cm}\\ 
{$^e$}Kavli Institute for Cosmology and Institute of Astronomy,\\ 
Madingley Road, 
Cambridge, CB3 0HA, UK.\vspace{.2cm}\\ 
{$^f$} INAF/Osservatorio Astronomico di Trieste,\\ 
Via Tiepolo 11, 34143, Trieste, Italy.\vspace{.2cm}\\ 
{$^g$} INFN-National Institute for Nuclear Physics\\
Via Valerio 2, 34127, Trieste, Italy.\vspace{.2cm}\\
}

\abstract{
We present forecasts for the accuracy of determining the parameters of a
minimal cosmological model and the total neutrino mass based on combined mock
data for a future {\it Euclid}-like galaxy survey and {\it Planck}. We consider
two different galaxy surveys: a spectroscopic redshift survey and a cosmic
shear survey. We make use of the Monte Carlo Markov Chains (MCMC) technique and
assume two sets of theoretical errors. The first error is meant to account for
uncertainties  in the modelling of the effect of  neutrinos on the non-linear
galaxy power spectrum and  we assume this error to be fully correlated in
Fourier space. The second error is meant to parametrize  the overall residual
uncertainties in  modelling the non-linear galaxy  power spectrum at small
scales, and is conservatively assumed to be uncorrelated and to increase with
the  ratio of a given scale to the scale of non-linearity. It hence increases
with wavenumber and decreases with redshift. With these two assumptions for the
errors and assuming further conservatively that the uncorrelated error rises
above  2\% at $k = 0.4 \, h/$Mpc and $z = 0.5$, we find that a future {\it
Euclid}-like  cosmic shear/galaxy survey achieves a 1-$\sigma$ error on
$M_{\nu}$ close to 32 meV/25 meV,  sufficient  for detecting the total neutrino
mass with good significance.  If the residual uncorrelated errors indeed rises
rapidly towards smaller scales in the non-linear regime as we have assumed here
then the data  on non-linear scales does not increase the sensitivity to the
total neutrino mass. Assuming instead a ten times smaller theoretical error
with the same scale dependence, the error on the total neutrino mass decreases
moderately from $\sigma(M_{\nu})$ = 18 meV to 14 meV when mildly non-linear
scales with $0.1\,h/$Mpc~$< k< 0.6\,h/$Mpc are included in the analysis of the
galaxy survey data.\\
}
   
\begin{document}

\section{Motivations}

Several ambitious ground-based and space-based galaxy surveys have
been planned for the next decade (e.g. {\sc
  ska}\footnote{\url{http://www.skatelescope.org/}}, {\sc
  lsst}\footnote{\url{http://www.lsst.org/lsst/}}), or are  about to take place (e.g. {\sc
  des}\footnote{\url{http://www.darkenergysurvey.org/}}). One of the
most ambitious approved missions, the {\sc {\it Euclid}}\footnote{\url{http://www.euclid-ec.org}} 
satellite \cite{Laureijs:2011mu}, is expected to be launched by ESA in 2019. It
will combine a galaxy redshift survey with weak lensing observations,
measuring the matter power spectrum and the growth of structure with unprecedented
accuracy. This will offer a unique opportunity to improve measurements
of cosmological parameters,  including the neutrino mass, known to slow down structure formation on 
intermediate and small scales\cite{Lesgourgues:2006nd}, as well as
constraints on dark energy and modified gravity models.

Recent constraints on the total neutrino mass appear to have converged on an upper limit of about
0.3 eV at the 95\% confidence level
(e.g. ~\cite{mantz010,reid010,thomas010,saito11,swanson012,signe012,Xia:2012na}), 
with the notable exception of Lyman$-\alpha$ forest data, which gives
an even lower bound of $0.17$eV \cite{seljak06}.
These constraints rely on a combination of data from Cosmic Microwave Background
(CMB) experiments such as WMAP, Baryonic Acoustic Oscillations (BAOs),
SuperNovae (SN) distance moduli, galaxy clustering and cosmic shear
(especially from the {\sc SDSS}\footnote{\url{http://www.sdss.org/}}
and {\sc CFHTLS}\footnote{\url{http://www.cfht.hawaii.edu/Science/CFHLS/}} surveys).
Data sets provided by Large Scale Structure (LSS) are
particularly important, since they are able to probe scales and
redshifts affected by neutrino free streaming both in the linear and
non-linear regimes.  Neutrino oscillation experiments provide a lower 
bound of $0.05$eV on the total neutrino mass, meaning that the 
allowed range is now significantly squeezed by cosmological data, 
and well within reach of future planned surveys.

Several forecasts have already been published on the sensitivity of
{\it Euclid} to cosmological parameters, with a focus on dark
energy, modified gravity, the neutrino mass, or other extensions
of the minimal $\Lambda$CDM model (see
e.g. \cite{Kitching:2008dp,Debono:2009bd,Wang:2010gq,Carbone:2010ik,Amendola:2011ie,Belloso:2011ms,Carbone:2011by,Hearin:2011bp,Hamann:2012mc}).
However reliable forecasts are difficult to obtain;
interpreting {\it Euclid} data on small (non-linear) scales will
require a more accurate modeling of systematic effects than is
currently achievable. This is true for both non-linear corrections to the
matter power spectrum, and for effects specific to each survey. 
In the case of the galaxy redshift survey, for instance, redshift space
distortions and scale-dependent bias. In the case of the cosmic shear survey, 
noise bias in shape measurements~\cite{Refregier:2012kg}. Some 
authors have pointed out that without considerable progress in modeling these effects, the
sensitivity to cosmological parameters might degrade considerably (see
e.g. \cite{Hearin:2011bp}).

Current forecasts tend either to incorporate only linear scales and
neglect these systematics, or to include a small range of mildly
non-linear scales and model systematics by including
nuisance parameters which are then marginalized over. Introducing such
nuisance parameters (for instance, in order to describe redshift-space
distortions) still assumes that we can predict the shape of these
effects, and reduce them to a simple family of curves. Hence, this
approach is not the most conservative.

On top of this, many forecasts are affected by a methodology issue:
apart from two recent works \cite{Hamann:2012mc,Wolz:2012sr}, they are
based on a Fisher matrix technique, whose results depend on the step
chosen in the calculation of numerical derivatives of the spectrum
with respect to the parameters (see
e.g. \cite{Wolz:2012sr,Perotto:2006rj}).

The present forecast has three objectives:

\begin{itemize}
\item
First, we wish to use a reliable forecast method for the sensitivity of a {\it Euclid}-like survey to $\Lambda$CDM parameters and to the total neutrino mass, 
based not on Fisher matrices, but on a parameter extraction from mock data with Markov Chain Monte Carlo (MCMC). This goal has also been achieved very 
recently by \cite{Hamann:2012mc}, although with a different approach for modeling the galaxy redshift survey. To our knowledge, the present analysis is 
the first MCMC forecast of a {\it Euclid}-like galaxy redshift survey using as an observable the power spectrum  $P(k)$ in wavenumber space. 

\item
Second, we wish to incorporate non-linear corrections using the most accurate available fitting formula accounting for neutrino mass effects, namely the version of {\sc halofit}~\cite{Smith:2002dz} presented in Ref.~\cite{Bird:2011rb}. This formula has been obtained by fitting to a suite of N-body simulations which incorporate neutrinos as free-streaming dark matter particles, using the code first presented in Ref.~\cite{viel010}.
The error in this formula specific to the neutrino mass was estimated by Ref.~\cite{Bird:2011rb} to be Gaussian, with squared variance
\begin{equation}
\alpha(k,z) \equiv \frac{\Delta P(k,z)}{P(k,z)} = \frac{\ln[1+k/k_\sigma(z)]}{1+\ln[1+k/k_\sigma(z)]} \, f_\nu~,
\label{alpha}
\end{equation}
where $f_\nu=\omega_\nu/\omega_m$ and $k_\sigma(z)$ is the non-linear wavenumber as defined and computed in {\sc halofit}. 
We include this in the likelihood as a fully correlated error, as described in detail in Appendix~\ref{A}, associated to a
unique nuisance parameter. 
\item
In order to obtain conservative results while keeping the analysis simple,
we will combine  this correlated error with  a second uncorrelated  error. 
This second uncorrelated error is assumed to 
account for extra uncertainties in our approximate modeling of non-linear corrections, redshift space distortions, scale-dependent bias and other 
systematic effects. By assuming an uncorrelated error on each data point, we remain more conservative than if we marginalized over a small set of 
nuisance parameters representing several types of fully correlated errors. Throughout this work, we assumed for convenience 
that the relative theoretical error on the power spectrum was given by Eq.~(\ref{alpha}), with $f_\nu$ replaced by a constant factor, by default $0.05$.
This error grows smoothly from zero on linear scales up  to 5\% on deeply non-linear scales. For a concordance cosmology and at redshift $z=0.5$, 
it reaches 1\% near $k=0.1\,h$Mpc$^{-1}$ and 2.3\% around $k=0.6\,h$Mpc$^{-1}$. We assume that ten years from now, this will provide 
a reasonable description of the total uncertainty coming from all systematic effects in each of the two surveys. Occasionally, we will consider the 
effect of dividing the magnitude of the error by two or ten, to evaluate the effect of better control of non-linear systematics. 
We emphasise that the exact form of the uncorrelated error is
obviously just an educated guess and that a different k-dependence will {\it e.g.}
influence the  assessment of how useful pushing to smaller scales will be.  
Of course, introducing a fully uncorrelated error (or alternatively, form filling functions as in  \cite{Kitching:2008eq}) 
is very conservative in that it assumes that no modeling of systematics is accurate enough. In several years from now, it might become 
realistic to model most systematics with several types of correlated errors, and to reduce the residual uncorrelated theoretical error 
to a smaller level than assumed in this work. 
\end{itemize}

\section{Galaxy redshift survey}

Throughout this paper, our fiducial model is chosen to be a flat $\Lambda$CDM model with three degenerate massive neutrino species. The fiducial parameter values are taken to be $\omega_b=0.02258$, $\omega_c=0.1109$, $A_s=2.43\times 10^{-9}$ (pivot scale $k_*=0.05\,h$Mpc$^{-1}$), $n_s=0.963$, $h=0.710$, $z_{\rm reio}=10.3$, $m_\nu=0.07$~eV (so $M_\nu=0.21$~eV). For the power spectrum of the mock data, we could take directly the fiducial power spectrum, or generate a random spectrum realization corresponding to the same model. As illustrated in~\cite{Perotto:2006rj}, the two options lead to the same forecast errors, so for simplicity we assume an observed power spectrum equal to the theoretical power spectrum of the fiducial model.

\FIGURE[h]{
\includegraphics[width=7.3cm]{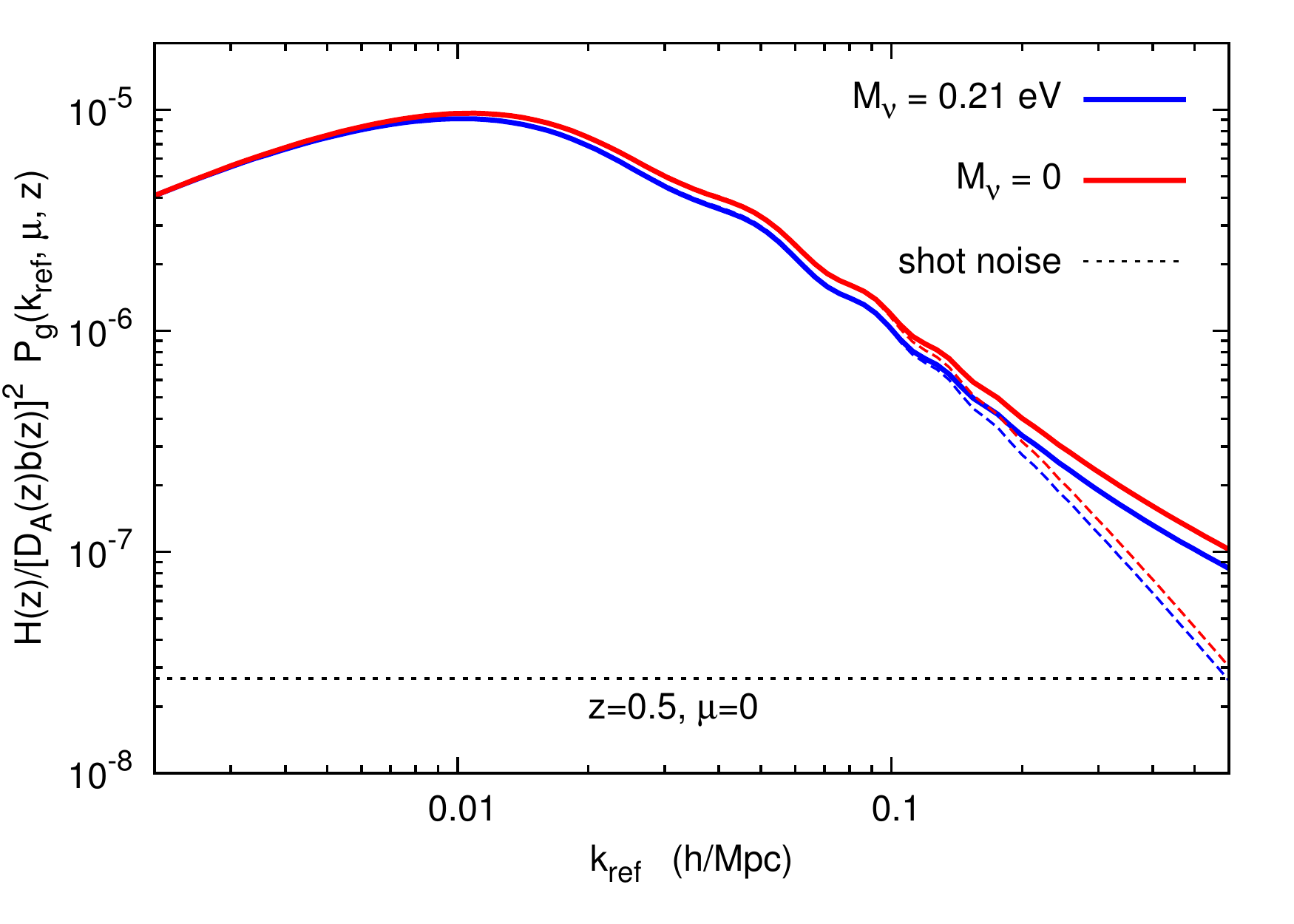}
\includegraphics[width=7.3cm]{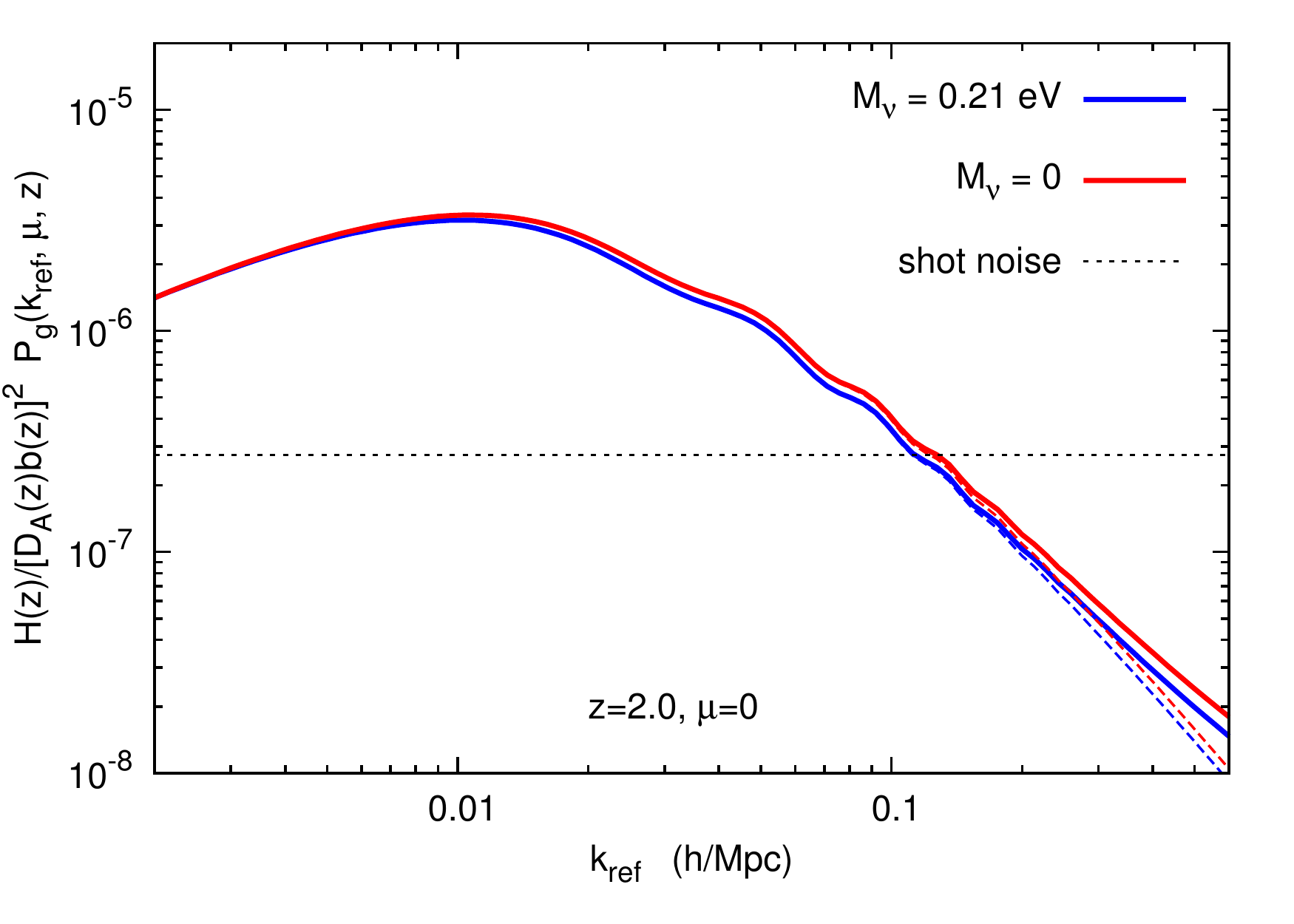}\\
\includegraphics[width=7.3cm]{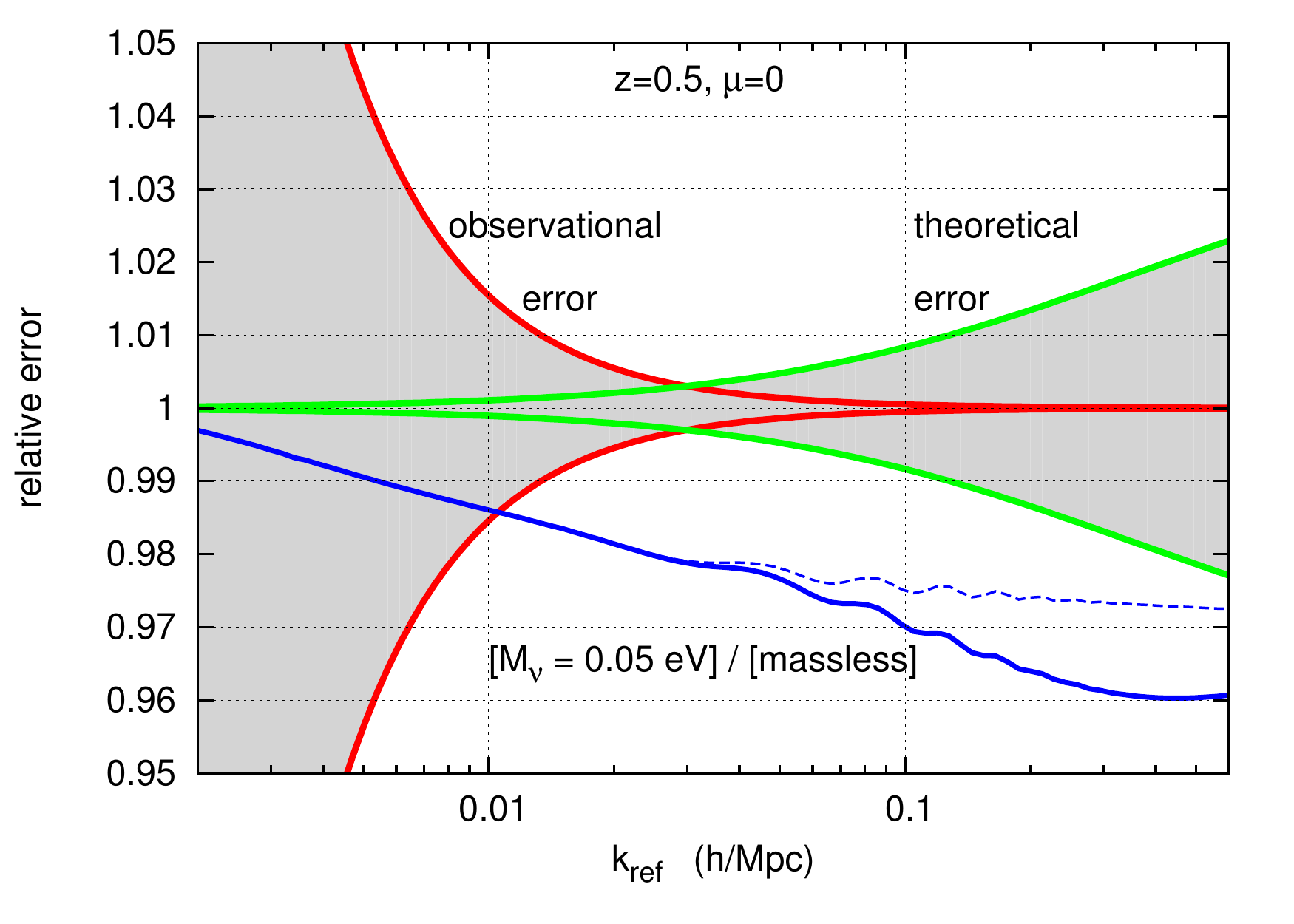}
\includegraphics[width=7.3cm]{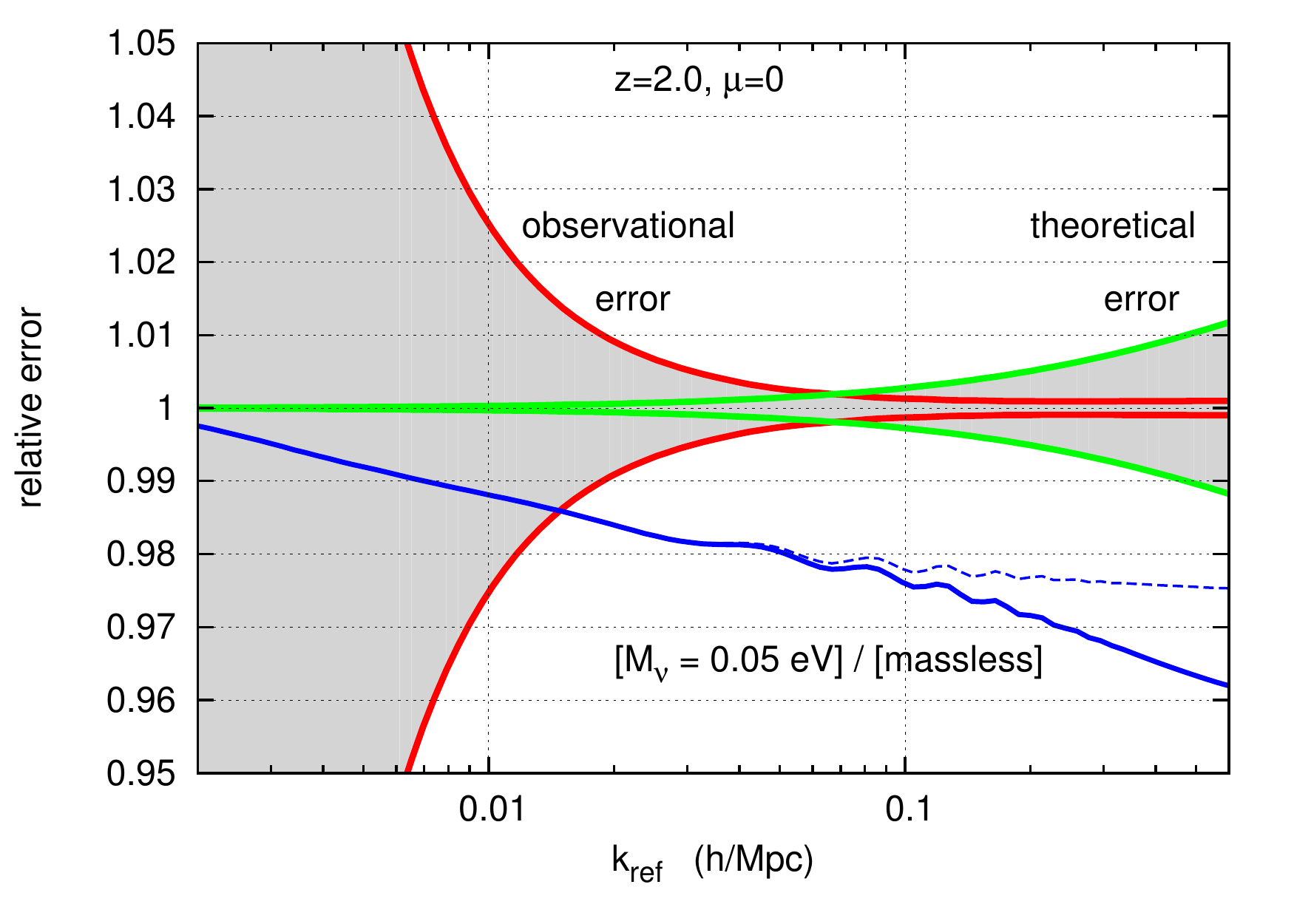}
\caption{Observable spectrum (top) and relative error on this spectrum
  (bottom), for the first redshift bin (left) and last redshift bin
  (right) of a {\it Euclid}-like galaxy redshift survey. The quantity
  displayed in the top is the galaxy power spectrum $P_g(k_{\rm ref}, \mu, z)$ as a function of the fiducial wavenumber $k_{\rm ref}$, for fixed redshift and perpendicularly to the line of sight ($\mu=0$), rescaled by the inverse squared bias $b(z)^{-2}$ and by a factor $H(z)/D_A(z)^2$: it is therefore a dimensionless quantity. The upper plots show a comparison between a model with massless neutrinos and our fiducial model ($M_\nu = 3 m_\nu = 0.21$~eV). Solid lines are derived from the non-linear matter power spectrum using the updated {\sc halofit} version of ref.~\cite{Bird:2011rb}, while  dashed lines are derived from the linear power spectrum. The lower plots show the part of the relative error coming from observational or theoretical errors only (cosmic variance is included in the observational error). In these plots, the individual 1-$\sigma$ error on each data point has been rescaled by the square root of the number of points, in such a way that the edges of the error bands correspond to a shift between theory and observation leading to $\Delta \chi^2=1$, when only the observational or theoretical error is incorporated in the likelihood expression. In these lower plots, we also show for comparison the ratio  between a massless model and a model with the minimum total mass allowed by neutrino experiments, $M_\nu=0.05$~eV.
}
\label{fig:pk}
}

We fit the mock and {\it Euclid}-like spectra using the MCMC code {\sc MontePython} \cite{montepython}. 
{\sc MontePython} uses the Metropolis-Hastings algorithm like {\sc CosmoMC} \cite{Lewis:2002ah}, but is interfaced
with {\sc class} \cite{Lesgourgues:2011re,Blas:2011rf} instead of {\sc camb} \cite{Lewis:1999bs}, is written in {\tt
python}, and has extra functionality; it will soon be released
publicly, including the {\it Euclid}-like likelihood codes used in this work.

Technical details of the assumed likelihood and our analysis are presented in Appendix~\ref{A}. Let us summarize here the essential points. As in most of the recent Fisher-matrix-based forecasts, we assume that the reduced data is described by a set of observable power spectra $P^{\rm obs}(k_{\rm ref}, \mu, z)$, related to the familiar non-linear matter power spectrum $P_{NL}(k,z)$ in a non-trivial way in order to take into account redshift space distortions, linear light-to-mass bias, spectroscopic redshift errors and the Alcock-Paczynsky effect (see \ref{A1}). Of course, this modeling is imperfect: for this reason we introduce a theoretical error. For instance, we do not take into account galactic feedback \cite{Semboloni:2011fe}, assuming that this contamination can be predicted by simulations up to the level of our residual theoretical error function. The arguments $k_{\rm ref}$ and $\mu$ of the observable power spectrum stand respectively for the observed wavenumber assuming the fiducial cosmology, and the cosine of the angle between the observed wavevector and the line of sight. We assume sixteen redshift bins with mean redshift ranging from 0.5 to 2, and bin widths of $\Delta z=0.1$. For a fixed theoretical model, each observed value of $P^{\rm obs}$ in a bin centered on the point $(k_{\rm ref}, \mu, z)$ follows, to a good approximation, a Gaussian distribution with variance
\begin{align}
\left(\Delta P^{\rm obs}\right)^2 = \frac{2 (2 \pi)^2 }{k_{\rm ref}^3 V_{\rm survey} d\mu [dk_{\rm ref}/k_{\rm ref}]} \left(P^{\rm th}+1/n_g\right)^2~,
\end{align}
where $d\mu$ is the size of the bins in $\mu$ space, and $[dk_{\rm ref}/k_{\rm ref}]$ the size of the logarithmic bins in wavenumber space (see \ref{A2}). The characteristics of the survey are encoded in $V_{\rm survey}$, the survey volume, and $n_g$, the comoving number density of galaxies accounting for shot noise (see \ref{A3}). Hence, if for {\it every} observed data point the theory and the observation differed by this amount, the effective $\chi^2$ would increase with respect to its minimum value by the number of data points, namely
\begin{align}
N = B \frac{2}{d \mu} \frac{\ln(k_{\rm max}/k_{\rm min})}{[dk_{\rm ref}/k_{\rm ref}]}~,
\end{align}
where $B$ is the number of redshift bins.

To illustrate this error, in figure~\ref{fig:pk}, we show the relative error bar on the observed spectrum in the first and last redshift bin, assuming no additional theoretical error. For the purpose of comparing with the theoretical error introduced below, we do not show as usual the error corresponding to a one-sigma deviation for each given data point; we divided each error by $\sqrt{N}$, in such a way that the edge of the error band corresponds to a deviation between the observed and theoretical spectrum leading to  $\Delta \chi^2=1$. Note that the displayed quantity $\pm \Delta P^{\rm obs}/(P^{\rm obs} \sqrt{N})$ does not depend on the width of the bins in ($k_{\rm ref}$, $\mu$, $z$) space, but only on $P^{\rm th}$, $V_{\rm survey}$ and $n_g$.

We incorporate the theoretical error in the likelihood in the way described in section \ref{A4}. In few words, this error is normalized in such a way that a shift between theory and observations by a relative amount $\alpha$ (the quantity defined in eq.~(\ref{alpha})) leads to an increase of the $\chi^2$ by one. This is achieved simply by adding a term $N (\alpha P^{\rm th})^2$ to the total error variance. 
Figure~\ref{fig:pk} shows the relative theoretical error on the observed spectrum, normalized in such a way that the edge of the error band corresponds to a deviation between the observed and theoretical spectrum leading to $\Delta \chi^2=1$ when the observational error is switched off. These edges are directly given by $\pm \alpha$.

We see in this figure that our assumption for $\alpha$ leads to an error of 1\% at $k=0.1h$Mpc$^{-1}$ and 2.5\% at $k=0.6h$Mpc$^{-1}$ for the first redshift bin centered on $z=0.5$. For the last redshift bin in the galaxy survey, centered on $z=2$, non-linear corrections appear on smaller scales, and the error is only 1\% at $k=0.6h$Mpc$^{-1}$.

\TABLE[h]{
\begin{tabular}{ccc|cccccccc}
$k_{\rm max}$ & un. & co. &  \multirow{2}*{$10^4\omega_b$} & \multirow{2}*{$10^4\omega_c$} & \multirow{2}*{$10^3n_s$} & \multirow{2}*{$10^{11}A_s$} & \multirow{2}*{$10^3h$} & \multirow{2}*{$z_{\rm reio}$} & {\small $3m_\nu=M_\nu$} \\
{\small $(h/\mathrm{Mpc})$} & err. & err. &  & &  &  &  &  & {\small (meV)} \\
\hline
0.1 & -- & -- &  1.2 &       6.2 & 2.8 & 3.0 & 4.1 & 0.38 & 18 \\
\hline
0.1 & {\scriptsize $1/10$} & -- &  1.2 & 6.9 & 2.8 & 3.1 & 4.5 & 0.39 & 18 \\
0.1 & {\scriptsize $1/2$} & -- &  1.3 &	 9.5 & 3.2 & 3.5 & 6.1 & 0.39 & 23 \\
0.1 & $\bullet$ & -- &  1.3 &	11 & 3.4 & 3.6 & 6.7 & 0.40 & 25 \\
0.1 & $\bullet$ & $\bullet$ &  1.3 &	11 & 3.4 & 3.6 & 6.7 & 0.40 & 25 \\
\hline
0.6 & -- & -- &  0.86 & 2.1 & 0.37 & 1.2 & 0.40 & 0.23 & 5.9 \\
\hline
0.6 & {\scriptsize $1/10$} & --    &   1.1 & 4.8 & 2.5 & 2.7 & 3.0 & 0.37 & 14 \\
0.6 & {\scriptsize $1/2$}   & --    &   1.2 & 8.6 & 3.2 & 3.4 & 5.7 & 0.39 & 22 \\
0.6 & $\bullet$                   & --    &  1.3 &10 & 3.4 & 3.6 & 6.7 & 0.39 & 25 \\
0.6 & $\bullet$ & $\bullet$ &  1.3 & 10 & 3.4 & 3.6 & 6.7 & 0.39 & 25 \\	
\end{tabular}

\caption{Marginalized 1-$\sigma$ error for each model parameter, in a fit of {\it Planck} + {\it Euclid}-like galaxy survey data. The different lines correspond to different choices of $k_{\rm max}$, to the inclusion or not of the global uncorrelated theoretical error (un. err.), divided by ten ({\scriptsize $1/10$}), by two ({\scriptsize $1/2$}), or full ($\bullet$), to that of the specific neutrino-related correlated error (co. err.), and to the use of the non-linear or linear power spectrum. The models with correlated error have one more nuisance parameter $e_\nu$ not shown here, with unit 1-$\sigma$ error.  \label{table:pk_results}}
}

We performed  several forecasts for a combination of {\it Planck} data and a {\it Euclid}-like galaxy redshift survey data. It should be stressed that the characteristics of {\it Euclid} are not yet finalized. Our choice for $V_{\rm survey}$ and $n_g(\bar{z})$, detailed in \ref{A3}, should be taken as indicative only. For {\it Planck}, we follow the method presented in \cite{Perotto:2006rj} and do not include lensing extraction. For the experimental {\it Planck} sensitivity, we use the numbers presented in the {\it Planck} Bluebook\footnote{\url{http://www.rssd.esa.int/SA/PLANCK/docs/Bluebook-ESA-SCI(2005)1\_V2.pdf}, page 4, Table 1.1 (using only the best three HFI channels: 100, 143 and 217 GHz).}. This is a rather conservative model since the sensitivities are based on 14 months of observations instead of 30. 

The differences between our forecasts reside in the maximum wavenumber, equal to $k_{\rm max}= 0.1$ or $0.6\,h$Mpc$^{-1}$, and in various prescription for the theoretical error: no error at all, the uncorrelated error described above and in \ref{A4} (divided by ten, by two, or full), or additionally the correlated error accounting for neutrino-mass-related effects (described in \ref{A5}). Since we are using an increasing theoretical error on non-linear scales, we expect the amount of information contained in the data to saturate above some value of $k_{\rm max}$: this is the reason we can consider such a high value as $0.6\,h$Mpc$^{-1}$.  We did not try even higher values, first because our result would not change, and second because our forecast would become unrealistic: deep in the non-linear regime, the Gaussian assumption for the likelihood breaks down.

Our results are presented in Table \ref{table:pk_results}. Parameters like $\omega_b$ and $z_{\rm reio}$ are well determined by CMB data, and their forecast error depends very mildly on our different assumptions. For other parameters, the redshift survey plays a crucial role in removing parameter degeneracies. In that case, even with $k_{\rm max}=0.1\,h$Mpc$^{-1}$, including the uncorrelated theoretical error makes a difference: the parameter sensitivity degrades by up to 70\% for $h$. The 68\% neutrino mass error bar degrades by 40\%, from $\sigma(M_\nu)=0.018$~eV to $\sigma(M_\nu)=0.025$~eV.

Assuming only this uncorrelated error, the cases $k_{\rm
  max}$=0.1$\,h$Mpc$^{-1}$ and $k_{\rm max}$=0.6$\,h$Mpc$^{-1}$ give
almost the same results. Hence, our assumption for the theoretical
error magnitude is such that most of the information is contained on
linear scales. Thanks to realistic (or at least conservative)
assumptions for the theoretical error, the results of our forecast are
nearly independent of the cut-off $k_{\rm max}$. Without a theoretical
error, increasing $k_{\rm max}$ to $0.6\,h$Mpc$^{-1}$  would lead to a
spectacular (but totally unrealistic) decrease of the error bars, with
$\sigma(M_\nu)=0.0059$~eV. 

If we are more optimistic and half the
uncorrelated error, the error bars decrease marginally, as can be seen in
the Table (lines starting with ``1/2''). The error on the neutrino
mass only decrease by $\sim 10$\%. Assuming no error at all implies
that the spectrum can be predicted up to the $0.1$\% level or better
on small scales. In comparison, assuming a precision of one percent is
not very different from assuming two percent. With the halved error,
the sensitivity to the neutrino mass increases from
$\sigma(M_\nu)=0.023$eV to $\sigma(M_\nu)=0.022$eV when including data
in the range from 0.1 to $0.6\,h$Mpc$^{-1}$. 

Finally, in a very
optimistic forecast with an error ten times smaller, we start to see
how extra information can be extracted from non-linear scales; the
error decreases from $\sigma(M_\nu)=0.018$~eV to
$\sigma(M_\nu)=0.014$~eV when pushing $k_{\rm max}$ from 0.1 to
$0.6\,h$Mpc$^{-1}$. 

The inclusion of an additional correlated error accounting for neutrino-mass-related systematics has a negligible impact on our results. In our forecast, the uncorrelated and correlated part of the error have similar amplitudes and the same shape; however the uncorrelated error allows much more freedom and thus leads significantly more conservative results: this explains why the correlated error has a comparatively small effect.
It should be stressed that our results depend not only on the assumed
error amplitude at a given scale and redshift, but also on the
wavenumber dependence of the error  function $\alpha$. Different
assumptions, with a steeper or smoother step  in the error function
around the scale of non-linearity, would lead to different forecasts. 
In particular, as already mentioned the actual benefit from
pushing to smaller, non-linear scales depends on the assumed
k-dependence of the residual uncorrelated theoretical error.

\FIGURE[h]{
\includegraphics[width=15cm]{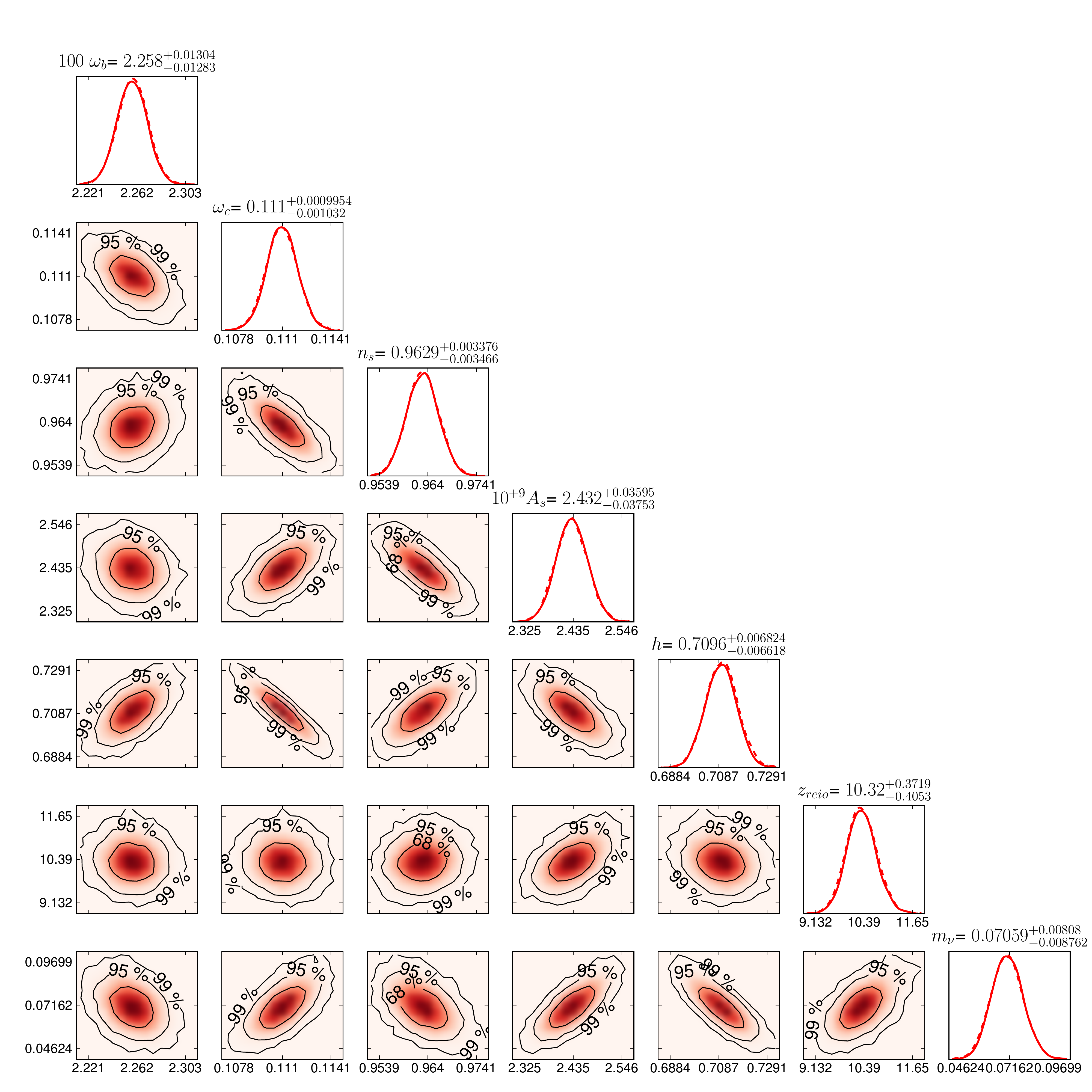}
\caption{Marginalized posteriors and two-dimensional probability contours in a fit of {\it Planck} plus a {\it Euclid}-like galaxy survey mock data, with $k_{\rm max}=0.6\,h$Mpc$^{-1}$ and a global uncorrelated theoretical error (second line starting from the bottom in Table~\ref{table:pk_results}).}
\label{fig:pk_triangle}
}


For the case with $k_{\rm max}=0.6\,h$Mpc$^{-1}$ and no neutrino-related correlated error, we show the one and two-dimensional posterior probability on cosmological parameters in figure~\ref{fig:pk_triangle}. We see several pronounced parameter degeneracies. For instance, the neutrino mass is very correlated with $\omega_c$ and $h$. This suggests that further progress could be made by including extra data sets, such as direct measurements of the Hubble parameter, the cluster mass function, supernovae luminosity, 21-cm anisotropies, and so on.

Our results are consistent with those of ~\cite{Carbone:2010ik,Carbone:2011by}, although a direct comparison is difficult, since those authors include two extra parameters, $w_0$ and $w_a$, in their forecast. The results in Table 2.1 of \cite{Amendola:2012ys}, based on the same cosmological model, match our prediction in the case with no non-linear scales and no theoretical error included. A similar sensitivity was found by~\cite{Hamann:2012mc} for a Euclid-like photometric redshift survey, referred to as ``cg'' in their Table~2. However, this reference presents other results based on even more conservative assumptions than ours. We assumed that the bias function for each redshift bin could be determined in advance (up to corrections on non-linear scales contained in our global theoretical error). This assumption has also been made in most recent forecasts, since both N-body simulations and higher-order statistics in the real data allow the prediction of the redshift-dependent bias of a given population of galaxies, at least on linear scales. Were this approach found to be unreliable, it would be necessary to marginalize over the linear bias in each redshift bin, $b(z_i)$. Ref.~\cite{Hamann:2012mc} did such a marginalization in the runs called ``cgb'' and ``cgbl'', with no prior at all on each $b(z_i)$. They found roughly the same error bar on $\omega_c$ and $h$ than in our forecast with theoretical error, but a much larger error on the neutrino mass. However, it seems unlikely that at the time when {\it Euclid} data will be analyzed, no information at all will be available on the linear bias of the observed population of galaxies.

\section{Cosmic shear survey}

For the case of a {\it Euclid}-like cosmic shear survey, we stick to
the same fiducial model and methodology as in the previous
section. The likelihood is now a function of the observed lensing
power spectrum $C_l^{{\rm obs}\,ij}$ in harmonic space and for each pair $ij$ of redshift bins, taking into account photometric redshift errors and shot noise (for details, see \ref{B1} and \ref{B2}). We assume experimental sensitivities summarized in \ref{B3}, and cut the observations in five redshift bins covering the range $0<z<3.5$ (although a negligible amount of galaxies contribute between 3 and 3.5). We do not take into account intrinsic alignment, assuming that this contamination can be removed up to the level of our residual theoretical error function \cite{Blazek:2012hm,Kirk:2011aw}.

As explained in detail in \ref{B4}, there is a small technical difference between the likelihood of the galaxy survey and the shear survey in the way we incorporate the uncorrelated theoretical error. For the galaxy survey, the theoretical error was encoded as an extra contribution to the total error variance. This can be justified mathematically by marginalizing over one nuisance parameter for each data point. The shape of the galaxy survey likelihood allows for an analytical minimization over each nuisance parameter, in such a way that nuisance parameters do not appear explicitly in the final likelihood. We found that no such scheme is accurate enough in the case of the (chi-square type) shear likelihood. Hence our likelihood routine performs an explicit minimization over one nuisance parameter per data point. For simplicity, we assume that the error is uncorrelated between different values of $l$, but not between different bins for a given $l$: this assumption could be relaxed, at the expense of increasing the computing time.

\FIGURE[h]{
\includegraphics[width=7.3cm]{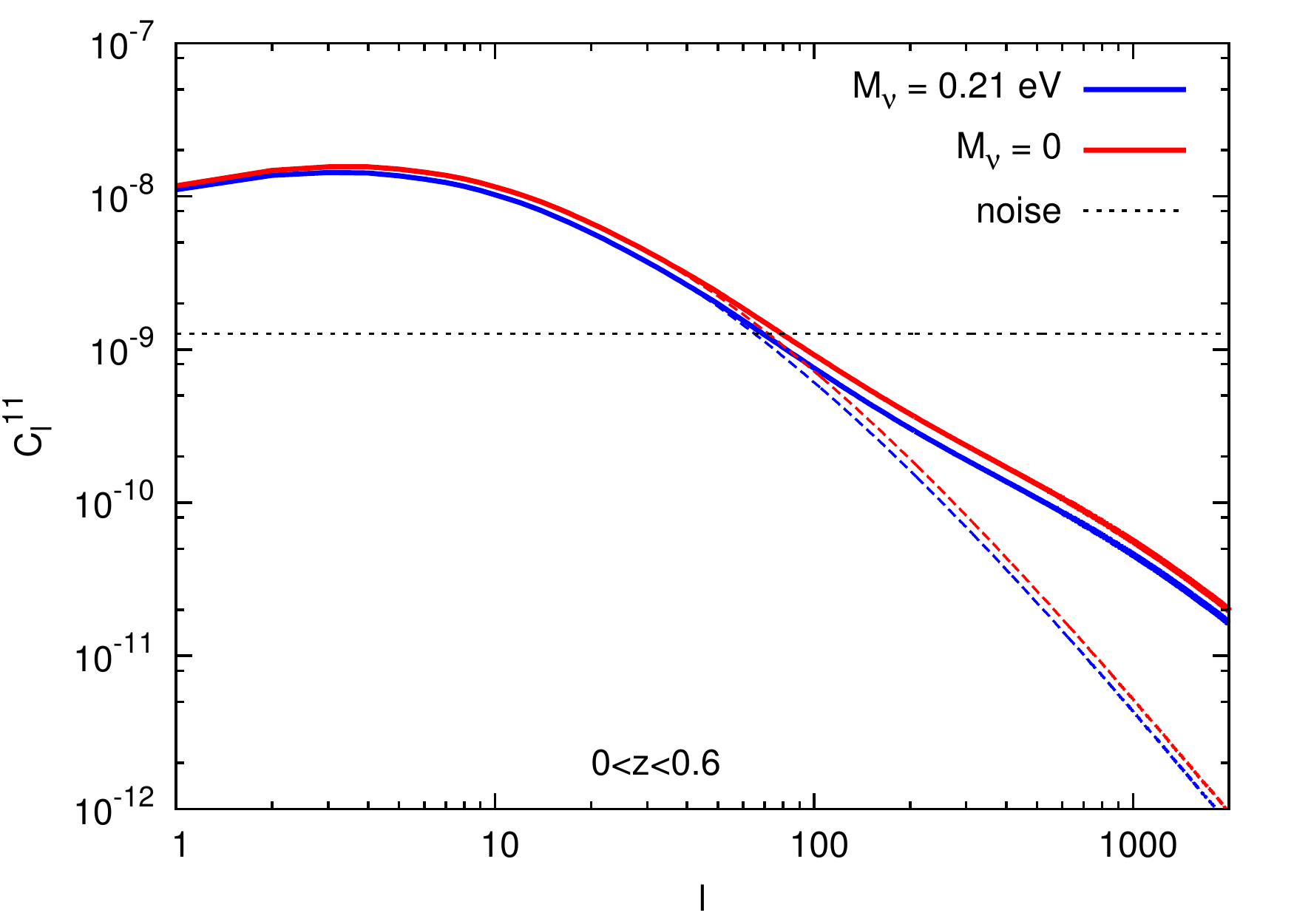}
\includegraphics[width=7.3cm]{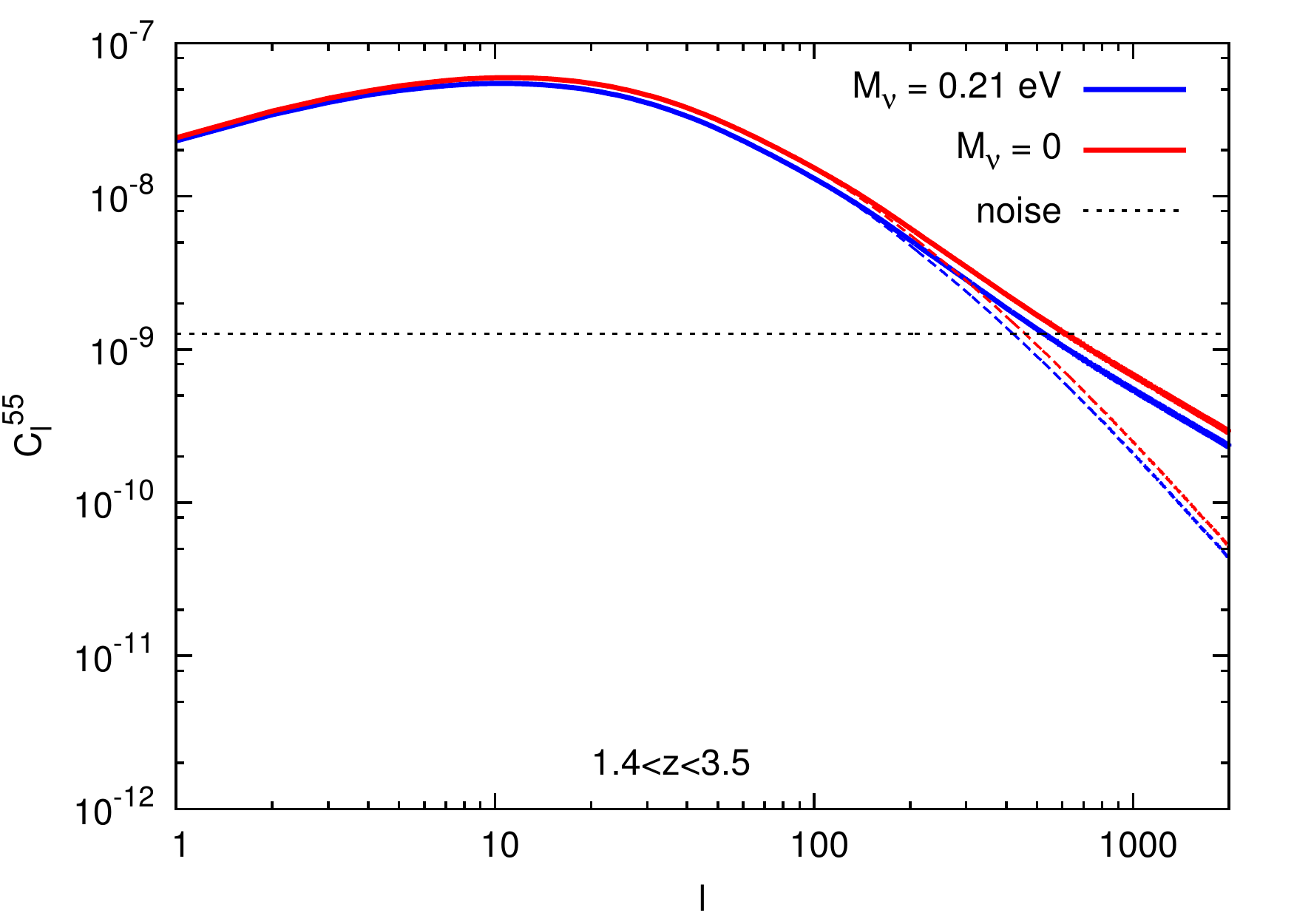}\\
\includegraphics[width=7.3cm]{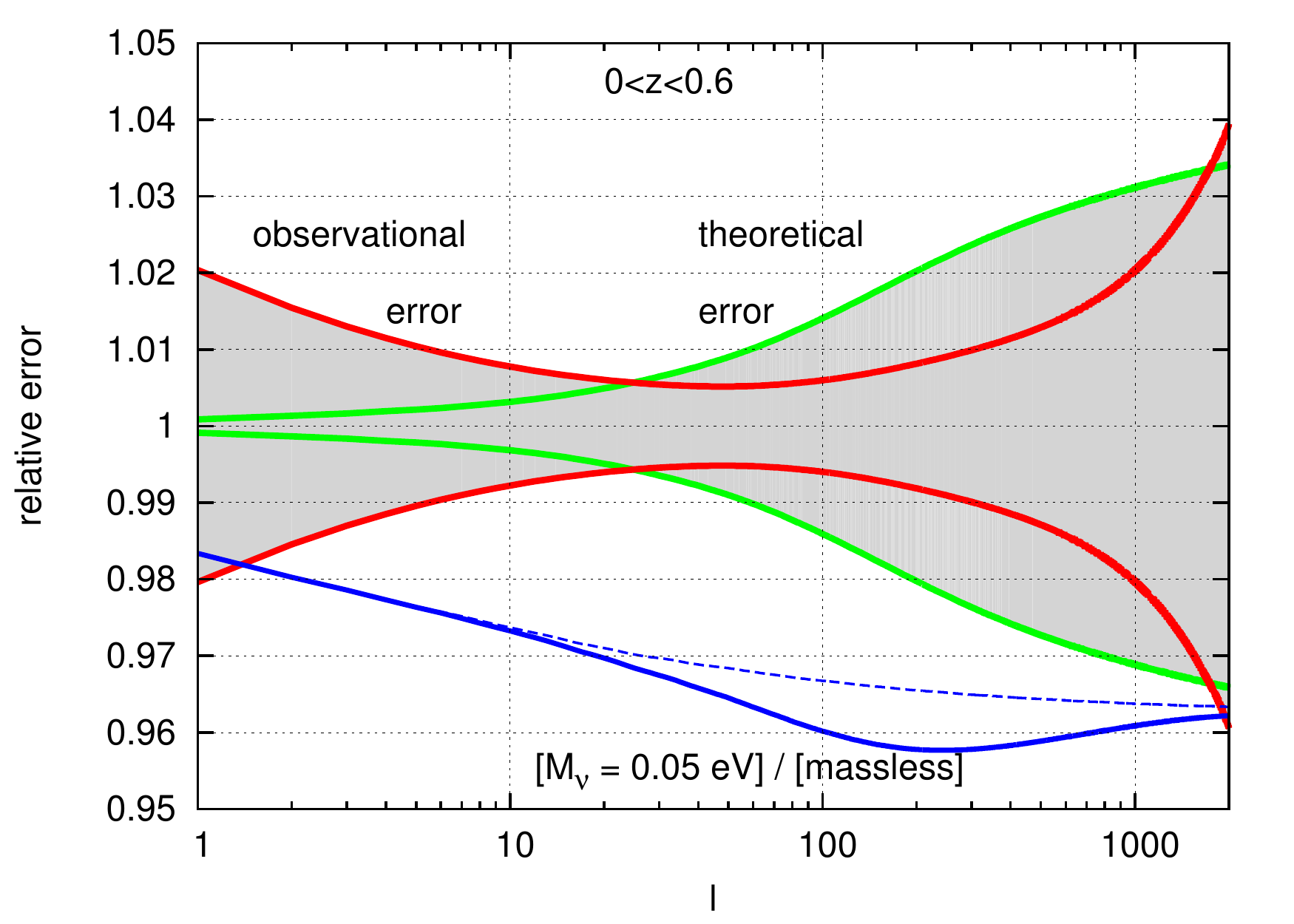}
\includegraphics[width=7.3cm]{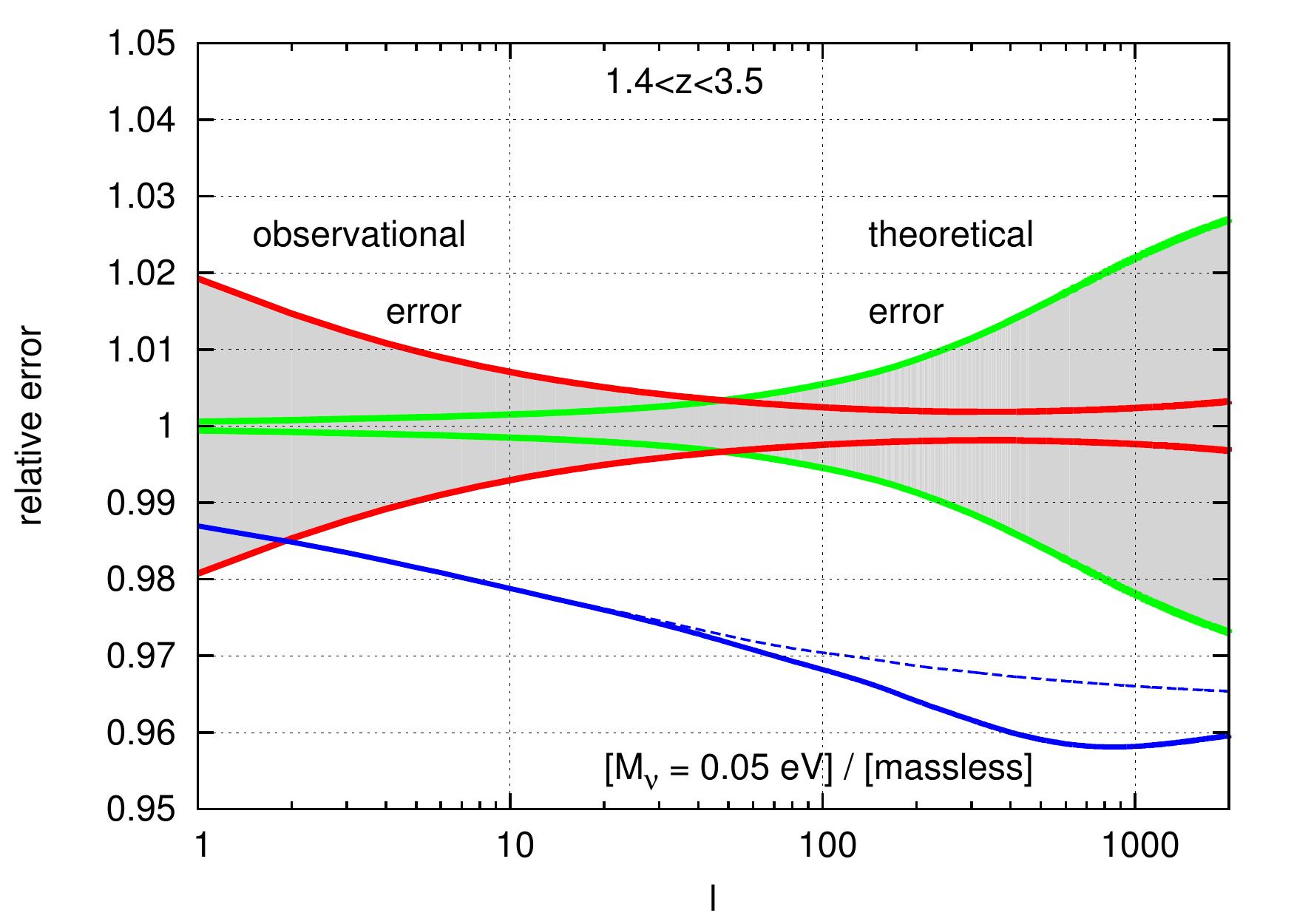}
\caption{
Observable cosmic shear power spectrum (top) and its relative error (bottom) for the first redshift bin (left) and last redshift bin (right) of a {\it Euclid}-like shear survey. The quantity displayed above is the lensing auto-correlation spectrum $C_l^{ii}$ (dimensionless). The upper plots show the comparison of a model with massless neutrinos to our fiducial model ($M_\nu = 3 m_\nu = 0.21$~eV). Solid lines are derived from the non-linear matter power spectrum using the recent update of {\sc halofit} \cite{Bird:2011rb}, while  dotted lines are derived from the linear power spectrum. The lower plots show the part of the relative error coming from observational or theoretical errors only (cosmic variance is included in the observational error). In these plots, the individual 1-$\sigma$ error on each data point has been rescaled by the square root of the number of points, in such a way that the edges of the error bands correspond to a shift between theory and observation leading to $\Delta \chi^2=1$, when only the observational or theoretical error is incorporated in the likelihood expression. In these lower plots, we also show for comparison the ratio  between a massless model and a model with the minimum total mass allowed by neutrino experiments, $M_\nu=0.05$~eV.
}
\label{fig:cs}
}

We fixed $l_{\rm max}=2000$, since beyond this value both the shot noise term and the theoretical error are large, as shown in figure~\ref{fig:cs}. This figure also shows the relative error on the observed spectrum in the first and last redshift bins, coming either from observational errors (including cosmic variance) or from the theoretical error, and using exactly the same conventions as in the previous section: the edges of each of the two error bands correspond to a shift between the theory and the observation leading to $\Delta \chi^2=1$ when either the observational or the theoretical error are included in the likelihood. The lowest redshift bin incorporates  small non-linear scales: this explains why at $l=2000$, the theoretical error reaches 3.5\%.

\TABLE[h]{
\begin{tabular}{cc|cccccccc}
un. & co. & \multirow{2}*{$10^4\omega_b$} & \multirow{2}*{$10^4\omega_c$} & \multirow{2}*{$10^3n_s$} & \multirow{2}*{$10^{11}A_s$} & \multirow{2}*{$10^3h$} & \multirow{2}*{$z_{\rm reio}$} & {\small $3m_\nu=M_\nu$} \\
err. & err. & & &  &  &  &  & {\small (meV)} \\
\hline
--              & --              & 1.1 & 3.9 & 2.4 & 2.8 & 4.0 & 3.7 &	26 \\
$\bullet$ & --              & 1.2 & 6.3 & 2.7 & 2.9 & 5.2 & 3.8 &	28 \\
$\bullet$ & $\bullet$ & 1.2 & 6.6 & 2.7 & 3.0 & 5.3 & 3.9 &	32 \\
\end{tabular}
\caption{
Marginalized 1-$\sigma$ error for each model parameter, in a fit to {\it Planck} + {\it Euclid}-like shear survey data. The different lines correspond to the inclusion or not of the global uncorrelated theoretical error (un. err.), and of the specific neutrino-related correlated error (co. err.). Our preferred prediction is given on the last line, and is very close to that of the second line. 
\label{table:cs_results}}
}

Our results are presented in Table~\ref{table:cs_results} for three
cases: no theoretical error, uncorrelated error only (described in
\ref{B4}), or additional neutrino-related correlated error (described
in \ref{B5}). The impact of the uncorrelated error is again important,
but not as pronounced as in the galaxy power spectrum case, because on
small scales the precision of the shear survey is limited by a
significant shot noise contribution. The neutrino mass error degrades
only from $\sigma(M_\nu)=0.026$~eV to 0.028~eV. For the shear survey we did not perform runs with a twice or ten times smaller error: the result for $\sigma(M_\nu)$ would simply lie between those two numbers. The impact of the neutrino-related error is small but further degrades the sensitivity to $\sigma(M_\nu)=0.032$~eV. While in the absence of theoretical error the galaxy survey seems more sensitive to the neutrino mass, the performance of the two methods are roughly identical once the same theoretical error ansatz is included.
 
\FIGURE[h]{
\includegraphics[width=15cm]{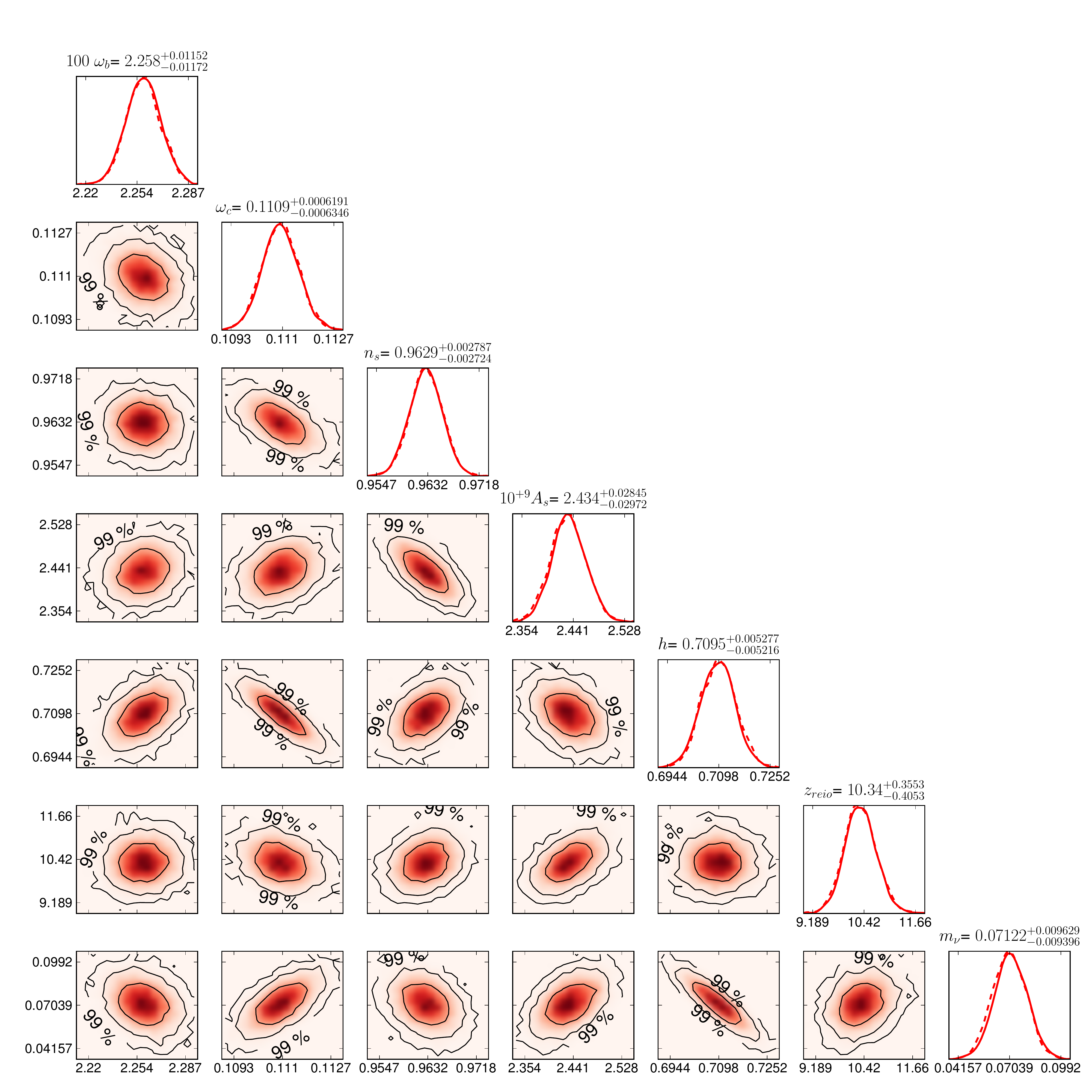}
\caption{Marginalized posteriors and two-dimensional probability contours in a fit of {\it Planck} + {\it Euclid}-like shear survey data, with a global uncorrelated error of 5\% on non-linear scales (second model in Table~\ref{table:cs_results}).
}
\label{fig:cs_triangle}
}

The triangle plot of figure~\ref{fig:cs_triangle} shows that the parameter degeneracies are very similar for the two cases of the galaxy survey and shear survey. Nevertheless, \cite{Hamann:2012mc} showed that combining the two data sets (with a proper cross-correlation matrix) leads to sensitivity improvements. It would be interesting to test this conclusion in presence of theoretical errors.

Our result are consistent with those of ~\cite{Kitching:2008dp}, although a direct comparison is difficult, since these authors include several extra parameters ($w_0$, $w_a$, $r$, $\alpha_s$) in their forecast. The predictions of~\cite{Hamann:2012mc} (case ``cs'' in their Table~2) lie between our results with and without theoretical errors. This is consistent since on the one hand, these authors use more optimistic survey characteristics ($d$, $\langle \gamma^2_{rms} \rangle$, $\sigma_{\rm ph}$), and on the other hand, we are including much larger values of $l$ (which is legitimate if our theoretical error is realistic).

\section{Conclusions}
We have presented forecasts of cosmological parameters by using, in combination 
with {\it Planck} data, two {\it Euclid}-like mock future data sets: a
galaxy spectroscopic redshift survey and a cosmic shear survey.  We focused our
attention on constraints that can be achieved on the total neutrino
mass by using the data in the linear and non-linear regimes.  

In order to do this conservatively we adopt the
following improvements with respect to similar works performed
recently in the literature: $i)$ we make use of Markov Chain Monte Carlo 
rather than the Fisher Matrix, which results in more
reliable error bars, as well as considering degeneracies between parameters. 
Ultimately, we found that the posterior probability is very close to a
multivariate Gaussian for the model considered. However, a Fisher matrix
approach could not have confirmed this, and would not have been
explicitly independent of the stepsize in the numerical derivatives.
$ii)$ we rely on a modification of {\sc HALOFIT} that accounts for 
massive neutrinos, and predicts the non-linear matter power spectrum to
small scales, based on the results of N-body and hydro
simulations.
$iii)$ we conservatively consider errors both on the
non-linear observable power at small scales and on the neutrino induced
suppression, and explictly show how to implement these errors in the
likelihood calculation.

It is instructive to see that with the shape assumed for the
uncorrelated theoretical error, and a conservative assumption on its
amplitude (leading to a 2\% error at $k_{\rm max}=0.4\,h$/Mpc and
$z=0.5$), the sensitivity to cosmological parameter is still
satisfactory. The error bar on the total neutrino mass, of the order
of 32~meV (cosmic shear) or 25~meV (redshift survey), would still allow for a two sigma
detection of the total neutrino mass in the minimal normal hierarchy
scenario. However, with this amplitude and k-dependence of the
theoretical error, essentially all the information comes from linear scales. The next interesting question is to check how much the uncorrelated error should be controlled in order to start being sensitive to mildly non-linear scales. Assuming a twice smaller error does not change the parameter sensitivity by a significant amount. Extracting significant information from non-linear scales requires an error ten times smaller, at the level of $0.2$\%. Here the error on the neutrino mass decreased from $\sigma(M_{\nu}) = 18$~meV to 14~meV when adding scales with $0.1<k<0.6\,h$/Mpc to the analysis. This shows that it would be extremely useful to be able to predict the observable power spectrum of a given cosmological model up to a residual uncorrelated error of the order of 0.1\% (resp. 0.2\%) at $k \sim$0.1$\,h$/Mpc (resp. $k \sim$0.4$\,h$/Mpc) and $z=0.5$. This will be a major challenge for theoretical and numerical cosmology in the next decade.

\section*{Acknowledgements}
We would like to thank Henk Hoekstra, Tom Kitching and Valeria Pettorino for their comments on this manuscript.
This project is supported by a research grant from the Swiss National
Science Foundation.  MV acknowledges support from grants: INFN/PD51,
ASI/AAE, PRIN MIUR, PRIN INAF 2009 and from the ERC Starting Grant
``cosmoIGM''.  BA and JL acknowledge support from the Swiss National Fundation.
SB is supported by NSF grant AST-0907969

\appendix

\section{Galaxy redshift survey implementation\label{A}}
\subsection{Observed spectrum\label{A1}}
Let $P^{\rm obs}$ be the observed/mock/fiducial power spectrum, and $P^{\rm th}$ the spectrum that one would expect to see given the theoretical model. Each of these quantities relates to the galaxy spectrum $P_g$ and finally to the total non-linear matter spectrum $P_{\rm NL}$ by taking into account redshift distortion effects,  spectroscopic redshift errors and light-to-mass bias. A good approximation of such a relation is given by (see e.g. \cite{Belloso:2011ms,Amendola:2011ie}):
\begin{align}
P^{\rm th/obs}(k_{ref \perp},k_{ref \parallel}, z) &=
\frac{D_A(z)^2_{\rm ref} H(z)}{D_A(z)^2 H(z)_{\rm ref}} 
P^{\rm th/obs}_g(k_{ref \perp},k_{ref \parallel}, z)~, \\
P^{\rm th/obs}_g(k_{ref \perp},k_{ref \parallel}, z) &= b(z)^2 \left[1+\beta(z,k)\frac{k^2_{ref \parallel}}{k^2_{ref \perp}+k^2_{ref \parallel}}\right]^2 P^{\rm th/obs}_{NL}(k,z)e^{-k^2 \mu^2 \sigma_r^2}~,
\end{align}
with the definitions
\begin{align}
\beta(k,z) &\equiv b(z)^{-1} \frac{d \ln[ P^{\rm th/obs}_{NL}(k,z) ]^{1/2}}{d \ln a} = \frac{1}{2 b(z)} \frac{d \ln P^{\rm th/obs}_{\rm NL}(k,z)}{d \ln a}~,\\
k_{{\rm ref} \perp} &= k_{\perp} H(z)_{\rm ref}/H(z), \quad
k_{{\rm ref} \parallel} = k_{\parallel} H(z)_{\rm ref}/H(z),\\
\mu & \equiv \hat{k}_{\rm ref}.\hat{r} =k_{{\rm ref} \parallel}/k_{\rm ref},\\
k^2 &= \left(\frac{(1-\mu^2) D_A(z)^2_{\rm ref}}{D_A(z)^2} 
+ \frac{\mu^2 H(z)^2}{H(z)^2_{\rm ref}}\right) k_{\rm ref}^2~.
\end{align}
Here $b(z)$ is the bias, assumed to be scale-independent in the range of scales of interest, $a$ is the scale factor, $H(z)$ is the Hubble parameter, $D_A(z)$ the angular diameter distance, and $\beta(z,k)$ accounts approximately for redshift space distortions.
So we can treat $k$ as a function of the arguments $(k_{\rm ref}, \mu, z)$ and write
\begin{align}
P^{\rm th/obs}(k_{\rm ref},\mu, z) &= \frac{D_A(z)^2_{\rm ref} H(z)}{D_A(z)^2 H(z)_{\rm ref}} b(z)^2 \left[1+\beta(z,k(k_{\rm ref},\mu,z))\mu^2\right]^2 \times \nonumber \\
& P_{\rm NL}^{\rm th/obs}(k(k_{\rm ref},\mu,z),z) e^{-k(k_{\rm ref},\mu,z)^2\mu^2\sigma_r^2} \label{def_P_th_obs}
\end{align}
\subsection{Likelihood\label{A2}}
For a narrow redshift bin $b$ centered on $\bar{z}$, the likelihood reads
\begin{align}
{\cal L}_b &= {\cal N}_b \exp \left[ 
-\frac{1}{2} \int_{k_{\rm min} < k_{\rm ref} < k_{\rm max}} \frac{d^3 \vec{k}_{\rm ref}}{(2 \pi)^3} V_{\rm eff}(k_{\rm ref},\mu,\bar{z}) \frac{(P^{\rm obs}(k_{\rm ref},\mu,\bar{z})-P^{\rm th}(k_{\rm ref},\mu,\bar{z}))^2}{2(P^{\rm th}(k_{\rm ref},\mu,\bar{z}))^2} \right] \\
&= {\cal N}_b \exp \left[ 
-\frac{1}{2} \int_{-1}^{1} d\mu \int_{k_{\rm min}}^{k_{\rm max}} \frac{k^2_{\rm ref} dk_{\rm ref}}{(2 \pi)^2} V_{\rm eff}(k_{\rm ref},\mu,\bar{z}) \frac{(P^{\rm obs}(k_{\rm ref},\mu,\bar{z})-P^{\rm th}(k_{\rm ref},\mu,\bar{z}))^2}{2(P^{\rm th}(k_{\rm ref},\mu,\bar{z}))^2} \right]~,
\end{align}
with an effective survey volume given by
\begin{equation}
V_{\rm eff}(k_{\rm ref},\mu,\bar{z}) = V_{\rm survey}(\bar{z}) 
\left[\frac{n_g(\bar{z})P_g^{\rm th}(k_{\rm ref},\mu,\bar{z})}{1+n_g(\bar{z})P_g^{\rm th}(k_{\rm ref},\mu,\bar{z})}\right]^2~.
\end{equation}
Later, we will specify the sensitivity of the survey, parameterized by $V_{\rm survey}$, $n_g$, $\sigma_r$, $k_{\rm min}$ and $k_{\rm max}$.
We skip here the derivation of the Fisher matrix, obtained by differentiating the above formula twice with respect to the cosmological parameters on which $P^{\rm th}$ depends, and evaluating this derivative at the maximum likelihood point. We checked that this calculation gives exactly the formula commonly used in the literature (see e.g. \cite{Belloso:2011ms,Amendola:2011ie}).
For the purpose of the discussion in the next section (and also of the numerical implementation), we wish to write explicitly the discrete limit of the integrals. We discretize $\mu$ in a set of equally spaced values $\mu_i$, and $l \equiv \ln k$ in a set of equally spaced values $l_j=\ln k_{{\rm ref} j}$. The step sizes are denoted $\Delta \mu$ and $\Delta l$ respectively. We then expand the integral as a sum, and for simplicity we omit the factors $1/2$ that should weight the boundary terms of each of the two integrals. We introduce the short-cut notations:
\begin{align}
N_{ij} &\equiv \Delta \mu \Delta l \frac{k_{{\rm ref} j}^3 V_{\rm eff}(k_{{\rm ref} j},\mu_i,z)}{(2 \pi^2)} ,\\
P^{\rm obs/th}_{ij} &\equiv P^{\rm obs/th}(k_{{\rm ref} j},\mu_i,z)  ,
\end{align}
and we get
\begin{align}
-2 \ln {\cal L}_b &= \sum_{i,j} \frac{\left(P^{\rm obs}_{ij}-P^{\rm th}_{ij}\right)^2}{2 (P^{\rm th}_{ij})^2/N_{ij}}.
\end{align}
This expression is easy to understand from first principles. Let us consider a single variable $\delta$ obeying a Gaussian distribution centered on zero and with variance $\langle \delta^2 \rangle = P$. If we observe $N$ independent realization $\delta_n$ of the variable $\delta$, we can build an estimator of the variance $P$ of $\delta$,
\begin{equation}
E = \frac{1}{N} \sum_n \delta_n^2.
\end{equation}
The variance of this estimator can be computed by noticing that each $\delta_n^2$ follows a $\chi^2$ distribution of order one, for which the mean is $P$ and the variance $2 P^2$. So  the sum $\sum_n \delta_n^2$ has a variance $2NP^2$. Finally $E$ has a variance $(2NP^2)/N^2=2P^2/N$. Moreover, $E$ is nearly Gaussian if $N$ is large, as a consequence of the central limit theorem. So the probability of the data $E$ given the theory $P$ is a Gaussian of mean $P$ and of variance $2P^2/N$. In other words,
\begin{equation}
-2 \ln {\cal L}(E|P) = \frac{(E-P)^2}{2P^2/N}~.
\end{equation}
The previous likelihood follows this form for each discrete term. Indeed each term corresponds to the likelihood of the estimator of the power spectrum in a thin shell in Fourier space. The number of independent measurements, i.e. of independent wavenumbers in each shell, is given by $N_{ij}$. The role of $E$ and $P$ is played respectively by $P^{\rm obs}_{ij}$ and $P^{\rm th}_{ij}$. Such a likelihood was first derived in pioneering papers like \cite{Feldman:1993ky,Tegmark:1997rp}.

\subsection{Survey specifications\label{A3}}

We computed this likelihood for values of $V_{\rm survey}(\bar{z})$, $n_g(\bar{z})$, $\sigma_r(\bar{z})$ inspired from currently plausible  {\it Euclid} specifications, which are likely to change over the next years.
We divide the observations into sixteen redshift bins of width $\Delta z=0.1$, ranging from $\bar{z}=0.5$ to $\bar{z}=2.0$. For each bin, we assumed: 
\begin{itemize}
\item a volume per bin $V_{\rm survey}(\bar{z}) = 4 \pi f_{\rm sky} [r(\bar{z})]^2 (1+\bar{z})^{-3}
\frac{\partial r(z)}{\partial z} \Delta z$, where $r(z)$ is the comoving distance
up to a comoving object with redshift $z$, with the explicit assumption that $a_0=1$:
\begin{equation}
r(z) = \int_0^{z} \frac{dz'}{H(z')}~.
\end{equation}
We assume a sky coverage $f_{\rm sky} = 0.375$.
\item
a galaxy number density per comoving volume $n_g(\bar{z})$, related to the number of galaxies per square degree $d_g(\bar{z})$ through
\begin{align}
n_g(\bar{z}) = \frac{d_g(\bar{z}) \times 41\,253\,{\rm deg}^2}{4 \pi [r(\bar{z})]^2 \frac{\partial r(z)}{\partial z} \,\Delta z}~.
\end{align}
For $d_g(\bar{z})$, we start from the number presented in Table~2 of \cite{Geach:2009tm} for the case of a limiting flux of $3\times 10^{-16}$erg\,s$^{-1}$cm$^{-2}$. Following the recommendation of that paper, we divide these numbers by $1.37$ in order to get conservative predictions. Finally, we multiply them by an efficiency factor  $\epsilon=0.25$ (standing for the redshift success rate). For instance, for the first redshift bin, this gives $d_g(\bar{z})=9376/1.37\times0.25=1710$~deg$^{-2}$.
\item
a spectroscopic redshift error $\sigma_r = \frac{\partial r(z)}{\partial z} \sigma_z$ with $\sigma_z=0.001(1+z)$.  
\item
a scale-independent linear bias $b(\bar{z})$. The choice of $b(\bar{z})$ values affects the final result less crucially than that of $d_g(\bar{z})$. We could adopt the predictions of \cite{Orsi:2009mj} inferred from N-body simulations, but for simplicity, our forecast is performed under the approximation $b(\bar{z})=\sqrt{1+\bar{z}}$. So, we assume in this forecast that the linear bias will be accurately measured or predicted for each bin, and that deviations from this prediction (coming  from the non-linear evolution) will be known up to the level described by the theoretical error function.
\item
$k_{\rm min}$ can be chosen arbitrarily close to zero without changing the results.
\item
we tested two values of $k_{\rm max}$:  $0.1$ and $0.6\,h$Mpc$^{-1}$.
\end{itemize} 

\subsection{Accounting for a global uncorrelated theoretical error\label{A4}}

To present a realistic forecast, one should model all the systematic
effects not accounted for by the previous likelihood formula, such as: theoretical
errors in the calculation of the linear and non-linear power spectrum,
scale-dependence of the bias on small scales, residual shot noise in galaxy
counts beyond the contribution already included in the definition of $V_{\rm
eff}$, residual errors in the modeling of redshift space distortion beyond the
above scheme. On top of these corrections, one may have to take into account
the fact that the likelihood is not Gaussian on strongly non-linear scales. In
this paper, we limit ourselves to mildly non-linear scales $k \leq k_{\rm
max}=0.6h$Mpc$^{-1}$, and assume that non-Gaussianity effects are sub-dominant
to the previously mentioned systematics. We also neglect to marginalize 
over residual shot noise in each redshift bin, because Ref.~\cite{Belloso:2011ms,Amendola:2011ie} 
found that this has a negligible impact. 


Understanding these various systematics is a major challenge for the future,
which should be addressed with better simulations and analytical modeling. Here
we want to keep the analysis simple, and model these systematic errors in a
simple way, by adding to the spectrum an uncorrelated theoretical error
function. By uncorrelated we mean that the errors made at different scales are
independent from each other, which is the most conservative possible
assumption. In this case, we can introduce an independent Gaussian-distributed 
nuisance parameter $\epsilon_{ij}$ for each data point, and marginalize over 
it -- or rather, to a very good approximation, minimize over it:
\begin{align}
-2 \ln {\cal L}_b &= \sum_{i,j} \min_{-\infty<\epsilon_{ij}<+\infty} \frac{\left[P^{\rm obs}_{ij}-\left(P^{\rm th}_{ij}+\epsilon_{ij} R_{ij}^{1/2} \right)\right]^2}{2 \left(P^{\rm th}_{ij}+\epsilon_{ij}R_{ij}^{1/2}\right)^2/N_{ij}}+\epsilon_{ij}^2,
\label{lik_pk_nuis}
\end{align}
where $R_{ij}$ is the theoretical error variance for a bin in $(\mu,k_{\rm
ref})$ space centered on $(\mu_i, k_{{\rm ref}j})$. 
As long as the theoretical error is assumed to be small, it is also a valid approximation to neglect the $\epsilon_{ij}$-dependence of the denominator, in order to find a simple analytic solution for $\epsilon_{ij}$, which, injected back in eq.~(\ref{lik_pk_nuis}), gives
\begin{align}
-2 \ln {\cal L}_b &= \sum_{i,j} \frac{\left(P^{\rm obs}_{ij}-P^{\rm th}_{ij}\right)^2}{2 (P^{\rm th}_{ij})^2/N_{ij}+R_{ij}}.
\label{eq_pk_dis}
\end{align}
In other words, the theoretical error variance simply adds up to the noise variance.

Note that we explicitly checked that it is legitimate to neglect the $\epsilon_{ij}$-dependence of the likelihood denominator when minimizing over $\epsilon_{ij}$. We also coded the full likelihood with explicit minimization over each $\epsilon_{ij}$, and found the same results up to very good accuracy.

We choose a numerical value of  $R_{ij}$ motivated mainly by the current
level of precision of the {\sc halofit} algorithm. We assume a relative error on the non-linear power spectrum of the form
\begin{align}
\alpha(k,z) \equiv \frac{\Delta P_{\rm NL}^{\rm th}(k,z)}{P_{\rm NL}^{\rm th}(k,z)} = \frac{\ln[1+k/k_\sigma(z)]}{1+\ln[1+k/k_\sigma(z)]} \, 0.05~,
\label{eqn:alpha_def}
\end{align}
where $k_\sigma(z)$ is the scale of non-linearity computed by {\sc halofit}.
This function increases from zero to 5\% around the scale of non-linearity. 
Using the function $k(k_{\rm ref},\mu,\bar{z})$, this error can easily be propagated to the theoretical observable spectrum
\begin{align}
\alpha(k_{{\rm ref}},\mu,\bar{z}) \equiv \alpha(k(k_{\rm ref},\mu,\bar{z}), \bar{z}) =  \frac{\Delta P^{\rm th}(k_{{\rm ref}},\mu,\bar{z})}{P^{\rm th}(k_{{\rm ref}},\mu,\bar{z})}~.
\end{align}
In terms of the discretized observable spectrum, the error reads 
\begin{align}
\alpha_{ij} = \alpha(k_{{\rm ref}\,j},\mu_i,\bar{z})~.
\end{align}
The error variance $R_{ij}$ should be proportional to the power spectrum variance $(\alpha_{ij} P_{ij}^{\rm th})^2$. We also 
assume that the error makes a constant contribution to each logarithmic
interval in the space where observations are performed, i.e. is of the form
\begin{align}
R_{ij} \propto (\alpha_{ij} P_{ij}^{\rm th})^2 \frac{k_{{\rm ref} j}}{d\mu \,\, dk_{\rm ref}}~.
\end{align}
We normalize the error variance $R_{ij}$ in such a way that a one-sigma theoretical error in each data point results in increasing the effective $\chi^2$ by one unit, namely,
\begin{align}
R_{ij} = \left[ 2 B \left( \ln \frac{k_{\rm max}}{k_{\rm min}}\right)\right] \, (\alpha_{ij} P_{ij}^{\rm th})^2 \frac{k_{{\rm ref} j}}{d\mu \,\, dk_{\rm ref}}~,
\end{align}
where $B$ is the number of bins.
The role of the normalization factor between squared brackets will become clear below.
%
The likelihood becomes (using eq.~(\ref{eq_pk_dis}) and going back to the continuous limit)
\begin{align}
{\cal L} &= \Pi_b \,\, {\cal N}_b \exp \left[ 
-\frac{1}{2} \int_{-1}^{1} \frac{d\mu}{2} \int_{k_{\rm min}}^{k_{\rm max}} 
\frac{dk_{\rm ref}}{k_{\rm ref}}
\frac{(P^{\rm obs}-P^{\rm th})^2}{
(P^{\rm th})^2
\left\{
\frac{
(2 \pi)^2}{k^3_{\rm ref} V_{\rm eff}} + \alpha^2 B \ln \frac{k_{\rm max}}{k_{\rm min}} \right\}} \right]~,
\end{align}
where we omitted the argument $(k_{\rm ref},\mu,\bar{z}_b)$ of the functions $P^{\rm th}$, $P^{\rm obs}$, $V_{\rm eff}$ and $\alpha$. 
If one assumes that the observed and theoretical spectra differ by $\alpha P^{\rm th}$ for each $(k,\mu,z)$, and that in the denominator the theoretical error dominates over the observational one ($V_{\rm eff}=\infty$), then
\begin{align}
{\cal L} &= \Pi_b \,\, {\cal N}_b \exp \left[ 
-\frac{1}{2} \int_{-1}^{1} \frac{d\mu}{2} \int_{k_{\rm min}}^{k_{\rm max}} \frac{dk_{\rm ref}}{k_{\rm ref}}\frac{1}{ B \ln \frac{k_{\rm max}}{k_{\rm min}}}\right] = \left( \Pi_b {\cal N}_b \right) \exp \left[ 
-\frac{1}{2} \right]~,
\end{align}
which corresponds to a shift by $\Delta \chi^2_{\rm eff}=1$ with respect to the maximum likelihood ${\cal L}= \Pi_b {\cal N}_b$. 

If we had assumed the error to be fully correlated, instead of increasing the denominator of the likelihood, we would have replaced $P^{\rm th}$ by $P^{\rm th}(1+\epsilon \alpha)$, multiplied the likelihood by $\sqrt{1/2\pi} \exp [- \epsilon^2/2]$, and marginalized/minimized over $\epsilon$. Then, the assumption $P^{\rm obs} = P^{\rm th}(1+\alpha)$ would correspond to an optimal choice $\epsilon=1$ in the large $V_{\rm eff}$ limit,
and would
also lead to a shift in $\Delta \chi^2_{\rm eff}$ by one unit with respect to the assumption $P^{\rm obs} = P^{\rm th}$. In our case, we obtain the same shifting while assuming statistically independent errors for each data point.

Finally, the likelihood can be simplified to
\begin{align}
{\cal L} &= \Pi_b {\cal N}_b \exp \left[ -\frac{1}{2} \int_{-1}^{1} \frac{d\mu}{2} \int_{k_{\rm min}}^{k_{\rm max}} \frac{dk_{\rm ref}}{k_{\rm ref}} \frac{\left(\frac{H_{\rm ref}}{D_{A{\rm ref}}^2}P^{\rm obs}_g-\frac{H}{D_A^2} P_g^{\rm th}\right)^2}{ \frac{(2 \pi)^2}{k^3_{\rm ref} V_{\rm survey}}  \left(\frac{H}{D_A^2}P^{\rm th}_g + \frac{H}{D_A^2}\frac{1}{n_g} \right)^2 + \left(\alpha \frac{H}{D_A^2} P_g^{\rm th}\right)^2 B \ln \frac{k_{\rm max}}{k_{\rm min}} } \right]~,
\end{align}
where we omitted the argument $\bar{z}_b$ in the functions $V_{\rm survey}$, $D_A$, $H$ and $n_g$. This is exactly the relation implemented in our code.

\subsection{Accounting for an extra neutrino-related error\label{A5}}

The impact of massive neutrinos on non-linear corrections to the power spectrum has been investigated in \cite{Bird:2011rb}. By comparing with N-body simulations including neutrino particles, the authors of \cite{Bird:2011rb} re-calibrated {\sc halofit}, with a new neutrino mass dependent correction. This fitting procedure is of course not perfect and adds a systematic error growing with the neutrino mass. It was found that the leading error can be described with a correction
\begin {equation}
P_{NL}(k)=P_{NL}^{\rm halofit}(k)(1+e_{\nu} \sigma_\nu(k,z)), \qquad
\sigma_\nu(k,z) = 
\frac{\ln[1+k/k_\sigma(z)]}{1+\ln[1+k/k_\sigma(z)]} f_\nu
\label{eq:e_nu}
\end{equation} 
with $f_\nu\equiv \omega_\nu / \omega_m$, and $e_\nu$ is an unknown correction of unit variance, that we will treat as a Gaussian nuisance parameter. Hence our final definition of the likelihood accounting for both types of error reads 
\begin{align}
{\cal L} &= {\cal N} \exp \left[ 
-\frac{1}{2} \int_{-1}^{1} \frac{d\mu}{2} \int_{k_{\rm min}}^{k_{\rm max}} 
\frac{dk_{\rm ref}}{k_{\rm ref}}
\frac{(P^{\rm obs}-[P^{\rm th}(1+e_\nu \sigma_\nu)] )^2}{
\left[P^{\rm th}(1+e_\nu \sigma_\nu)\right]^2
\left[
\frac{
(2\pi)^2}{k^3_{\rm ref} V_{\rm eff}} + \alpha^2 B \ln{\frac{k_{\rm max}}{k_{\rm min}}}\right]} \right] \nonumber \\
&\times \frac{1}{\sqrt{2 \pi}} \exp\left[-\frac{1}{2} e_\nu^2\right]~,
\end{align}
where we omitted the argument $(k_{\rm ref},\mu,\bar{z})$ of the functions $P^{\rm obs}$, $P^{\rm th}$, $\sigma_\nu$, $\alpha_\nu$ and $V_{\rm eff}$. Note that the correction proportional to $e_\nu$ should not be added to $P^{\rm obs}$ since we are assuming for simplicity that the fiducial value of $e_\nu$ in the mock data is zero.

\section{Cosmic shear survey implementation\label{B}}
\subsection{Observed spectrum\label{B1}}
As in e.g. \cite{Belloso:2011ms,Amendola:2011ie},  we define the likelihood of the shear auto  or cross-correlation power spectrum in bins $i$ and $j$:
\begin{equation}
C_l^{ij} = H_0^4 \int_0^\infty \frac{dz}{H(z)} \, W_i(z) \, W_j(z) \, P_{\rm NL} \left (k=\frac{l}{r(z)}, z\right)~.
\end{equation}
Here, $W_i(z)$ is the window function of the $i$'th bin. It can be evaluated as a function of the radial distribution of galaxies in each redshift bin, $D_i(z)$, obtained by convolving the full radial distribution $D(z)$ with the photometric redshift uncertainty function ${\cal P}(z,z_{\rm ph})$, multiplied the top-hat window function of each bin:
\begin{eqnarray}
W_i(z)&=&\frac{3}{2} \Omega_m (1+z) F_i(z)\\
F_i(z)&=&\int_0^{\infty} \frac{n_i(z_s) (r(z_s)-r(z))}{r(z_s)} dz_s\\
n_i(z) &=& \frac{D_i(z)}{\int_0^\infty D_i(z') dz'}\\
D_i(z) &=& \int_{z_i^{\rm min}}^{z_i^{\rm max}}  {\cal P}(z,z_{\rm ph}) \, D(z_{\rm ph}) \, dz_{\rm ph}~.
\end{eqnarray}
The radial distribution $D(z)$ can be arbitrarily normalized, since $n_i(z)$ is anyway normalized to one.
We will assume that the photometric redshift uncertainty function is normalized to $\int_0^\infty {\cal P}(z,z_{\rm ph}) dz_{\rm ph}=1$, but a different normalization would not impact the final result for the same reason as for $D(z)$.
The noise spectrum contaminating the measurement of $C_l^{ij}$ is given by the diagonal matrix in ${ij}$ space:
\begin{equation}
N_l^{ij} = \delta_{ij} \langle \gamma_{rms}^2 \rangle n_i^{-1} ~,
\end{equation}
where $\langle \gamma_{rms}^2 \rangle^{1/2}$ is the root mean square intrinsic shear (like in the forecasts of the Euclid Red Book \cite{Laureijs:2011mu}, we assume that this quantity is equal to 0.30), and $n_i$ is the number of galaxies per steradian in the $i$'th bin, given by
\begin{equation}
n_i = 3600 \, d \, (180/\pi)^2 \hat{n}_i~,
\end{equation}
where $d$ is the full number of galaxies per square arcminute in all bins, and $\hat{n}_i$ is the fraction of galaxies in the $i$'th bin, given by:
\begin{equation}
\hat{n}_i= \frac{\int_{z_i^{min}}^{z_i^{max}} D(z)}{ \int_0^{\infty} D(z)}~.
\end{equation} 
We used the survey specifications for $D(z)$, ${\cal P}(z)$, $d$ and $f_{sky}$ detailed in Appendix~\ref{B3}.

Using $dz/dr=H$, we can write the same integrals in a different way (used in other papers and in our code):
\begin{align}
  C_l^{ij} = \frac{9}{16} \Omega_m^2 H_0^4 \int_0^\infty dr \, r^{-2} g_i(r) \, g_j(r) \, P \left (k=\frac{l}{r}, z(r)\right)\label{eqn:shear_pk_def}
\end{align}
with
\begin{eqnarray}
g_i(r) &=& 2 r (1+z(r)) \int_0^\infty dr_s \frac{\eta_i(r_s) (r_s-r)}{r_s}\\
\eta_i(r) &=& H(r) n_i(z(r))
\end{eqnarray}
and $n_i(z)$ is the same as before. 

\subsection{Likelihood\label{B2}}
 Let's assume some theoretical spectra $C_l^{\rm th\,ij}$ (here, the spectra of each model that we want to fit to the data, exploring the space of free cosmological parameters), and some observed spectra $\tilde{C}_l^{\rm obs\,ij}$. The matrix $\tilde{\bf C}^{\rm obs}_l$ of element $\tilde{C}_l^{\rm obs\,ij}$ is called the data covariance matrix. 
It can be inferred from the observed multipoles $a_{lm}^{{\rm obs} \, i}$, which are Gaussian distributed with a variance independent of $m$ in an ideal full-sky experiment, so that
\begin{equation}
\tilde{C}_l^{\rm obs\,ij}=(2l+1)^{-1} \sum_{m=-l}^l [a_{lm}^{\rm obs\,i*} a_{lm}^{\rm obs\,j} ]~.
\end{equation}
For a parameter forecast, instead of the covariance matrix of mock data, we can use some fiducial spectra corrected by the noise spectra of the experiment at hand:
\begin{equation}
\tilde{C}_l^{\rm obs\,ij} = C_l^{\rm fiducial\,ij} + N_l^{ij}~.
\end{equation}
This data covariance matrix should be compared with the theoretical covariance matrix defined as
\begin{equation}
\tilde{C}_l^{\rm th\,ij} = C_l^{\rm th\,ij} + N_l^{ij}~.
\end{equation}
We define the determinant of these $N \times N$ symmetric matrices:
\begin{eqnarray}
d_l^{\rm th} &=& \det \left( \tilde{C}_l^{\rm th\,ij}  \right) \\
d_l^{\rm obs} &=& \det \left( \tilde{C}_l^{\rm obs\,ij}  \right)~.
\end{eqnarray}
The determinants are homogeneous polynomials of order $N$ in the spectra, e.g. for $N=2$:
\begin{equation}
d_l^{\rm th} = \tilde{C}_l^{\rm th \, 11} \tilde{C}_l^{\rm th \, 22} - \, (\tilde{C}_l^{\rm th \, 12})^2~.
\end{equation}
The quantity $d_l^{\rm mix}$ can be built starting from $d_l^{\rm th}$, and replacing one after each other the theoretical spectra $\tilde{C}_l^{{\rm th} \, ij}$ by the corresponding $\tilde{C}_l^{{\rm obs} \, ij}$, e.g. for $N=2$:
\begin{equation}
d_l^{\rm mix} = \tilde{C}_l^{\rm obs \, 11} \tilde{C}_l^{\rm th \, 22} + \tilde{C}_l^{\rm th \, 11} \tilde{C}_l^{\rm obs \, 22} - 2 \, \tilde{C}_l^{\rm th \, 12} \tilde{C}_l^{\rm obs \, 12}~.
\label{def_dmix}
\end{equation}
So, $d_l^{\rm mix}$ is always linear in the $\tilde{C}_l^{{\rm obs} \, ij}$'s.
By construction, when $\tilde{C}_l^{\rm th\,ij}=\tilde{C}_l^{\rm obs\,ij}$, one has $d_l^{mix} = N d_l^{\rm th} = N d_l^{\rm obs}$. Since in an ideal full-sky experiment, the different multipoles are uncorrelated in $(l,m)$ space, the 
likelihood of the observed spectra given the theoretical spectra is as simple as:
\begin{equation}
{\cal L} = {\cal N} \, \Pi_{lm} \left\{ \frac{1}{(d_l^{\rm th})^{1/2}}\exp\left[ - \frac{1}{2} {\bf a}_{lm}^{\rm obs \, \dagger}  (\tilde{\bf C}^{\rm th}_l)^{-1} {\bf a}_{lm}^{\rm obs} \right] \right\}~,
\end{equation}
where ${\bf a}^{\rm obs}_{lm}=\left\{a_{lm}^{\rm obs\,i} \right\}$ is the N-dimensional vector of observed multipoles in each bin,  $\tilde{\bf C}^{\rm th}_l$ is the theoretical covariance matrix of element
$C^{\rm th\,ij}_{l}$ and ${\cal N}$ is a normalisation factor. After some simple algebra\footnote{in particular, using $A^{-1}={\rm adj}(A) / \det(A)$ where ${\rm adj}(A)$ is the adjugate matrix of $A$, i.e. the transpose of the matrix of cofactors of $A$.}, the likelihood simplifies to
\begin{equation}
{\cal L} = {\cal N} \, \Pi_{l} \left\{ \frac{1}{(d_l^{\rm th})^{1/2}} \exp\left[ - \frac{(2l+1)}{2} \frac{d_l^{mix}}{d_l^{\rm th}} \right] \right\}~.
\end{equation}
The effective chi square
\begin{equation}
\chi^2_{\rm eff} \equiv -2 \ln {\cal L} = - 2 \ln {\cal N} + \sum_l (2l+1) \left( \frac{d_l^{mix}}{d_l^{\rm th}} + \ln
d_l^{\rm th} \right)~,
\end{equation}
reaches its minimum for $\tilde{\bf C}_l^{obs}=\tilde{\bf C}_l^{th}$, corresponding to
\begin{equation}
\chi^{2\, \rm min}_{\rm eff} \equiv -2 \ln {\cal L}_{\rm max}  = -2 \ln {\cal N} + \sum_l (2l+1) \left( N + \ln
d_l^{\rm obs} \right)~.
\end{equation}
The $\chi^2$ relative to the best-fit model is then equal to
\begin{equation}
\Delta \chi^2_{\rm eff} \equiv -2 \ln \frac{{\cal L}}{{\cal L}_{\rm max}} = \sum_l (2l+1) \left( \frac{d_l^{mix}}{d_l^{\rm th}} + \ln
\frac{d_l^{\rm th}}{d_l^{\rm obs}} - N \right)~.
\end{equation}
Finally, a first-order approximation to account for the limited sky coverage of a given experiment, consists of increasing the cosmic variance by a factor $f_{\rm sky}^{-1/2}$, equivalent to postulating:
\begin{equation}
\Delta \chi^2_{\rm eff} \equiv \sum_l (2l+1) f_{\rm sky} \left( \frac{d_l^{mix}}{d_l^{\rm th}} + \ln
\frac{d_l^{\rm th}}{d_l^{\rm obs}} - N \right)~.
\end{equation}
This is precisely the expression used in the code.

\subsection{Survey specifications\label{B3}}

A given survey is specified by $D(z)$, ${\cal P}(z)$, $d$, and finally by the covered faction of the sky $f_{sky}$; it can then be decomposed in redshift bins according to some strategy defined by the user. For a {\it Euclid}-like experiment we use the same characteristics as in the Euclid Red Book \cite{Laureijs:2011mu}:
\begin{eqnarray}
D(z) &=& z^2 \exp [ - (z/z_0)^{1.5} ]  \quad {\rm for} \quad z<z^{max}=3.5 \\
&& {\rm with~mean~redshift} \quad z_{\rm mean}=1.412 z_0 = 0.9 \nonumber \\
{\cal P}(z,z_{\rm ph}) &=& \frac{1}{\sqrt{2 \pi \sigma_{\rm ph}^2}} \exp\left[-\frac{1}{2}\left(\frac{z-z_{\rm ph}}{\sigma_{\rm ph}}\right)^2\right]  \\
&&{\rm with} \quad \sigma_{\rm ph} = 0.05(1+z) \nonumber \\
d &=& 30 \, {\rm arcmn}^{-2}\\
f_{sky}&=&0.375~.
\end{eqnarray}
We assume five bins, with the first bin starting at $z_1^{min}=0$, the last one ending at $z_N^{max}=3.5$, and bin edges 
$z_i^{min} = z_{i-1}^{max}$ chosen such that each bin contains the same number of galaxies, i.e. $\hat{n}_i=1/N$. 

\subsection{Accounting for a global uncorrelated theoretical error\label{B4}}

Like for the power spectrum likelihood, taking into account an uncorrelated error on each data point is equivalent to minimizing over a number $L \equiv (l_{\rm max}-l_{\rm min}+1)$ of nuisance parameters $\epsilon_l$:
\begin{equation}
\Delta \chi^2_{\rm eff} \equiv \sum_{l=l_{\rm min}}^{l_{\rm max}} \min_{-\infty<\epsilon<+\infty} \left[ (2l+1) f_{\rm sky} \left( \frac{\tilde{d}_l^{mix}(\epsilon_l)}{\tilde{d}_l^{\rm th}(\epsilon_l)} + \ln
\frac{\tilde{d}_l^{\rm th}(\epsilon_l)}{d_l^{\rm obs}} - N\right)  + \epsilon_l^2\right]~.
\end{equation}
Here, $\tilde{d}_l^{\rm th}(\epsilon_l)$ stands for the determinant of the theory covariance matrix shifted by the theoretical error covariance matrix $R^{ij}_l$:
\begin{align}
\tilde{d}_l^{\rm th}(\epsilon_l) = \det( \tilde{C}_l^{\rm th\,ij} + \epsilon_l R^{ij}_l ).
\end{align}
Similarily, $\tilde{d}_l^{\rm mix}(\epsilon_l)$ stands for the sum of $N$ terms, each one being the determinant of  a matrix built from $ \tilde{C}_l^{\rm th\,ij} + \epsilon_l R^{ij}_l$, where one column has been replaced by the same column in the observed covariance matrix. Hence the quantity ${d}_l^{\rm mix}$ defined just above eq.~(\ref{def_dmix}) is identical to $\tilde{d}_l^{\rm mix}(0)$.

Note that for simplicity, we consider here uncorrelated errors for each $l$, but not for each bin. This approach could easily be generalized to independent bin errors, at the expense of introducing more nuisance parameters.

In the case of the power spectrum likelihood, we could find an analytical approximation of the  nuisance parameter value minimizing the effective $\chi^2$. In the present case, we checked that simple approximate solutions are not accurate enough. We perform a numerical minimization over each $\epsilon_l$ within the likelihood routine, using Newton's method.

We define our theoretical error covariance matrix $R^{ij}_l $ in a similar way as for the power spectrum likelihood. We start from the power spectrum relative error function $\alpha(k,z)$ defined in eq.(\ref{eqn:alpha_def}). The power spectrum error can be propagated to a covariance matrix error $E_l^{ij}$:
\begin{align}
E_l^{ij} = \frac{9}{16} \Omega_m^2 H_0^4 \int_0^\infty dr \, r^{-2} g_i(r) \, g_j(r) \,\alpha \left(k=\frac{l}{r},z(r)\right) P^{\rm th} \left (k=\frac{l}{r}, z(r)\right)~.\label{eqn:shear_E_l_def}
\end{align}
The theoretical error matrix $R_l^{ij}$ should be proportional to $ E_l^{ij}$. We normalize it to 
\begin{equation}
R_l^{ij} = L^{1/2} E_l^{ij},
\end{equation}
in such a way that enforcing a one-sigma theoretical error for each $l$ results in an increase of the $\chi^2$ by one (as would be the case for a fully correlated theoretical error with the same amplitude). Then, if one assumes that for each $l$ the observed spectra are equal to the theoretical ones shifted by a one-sigma theoretical error ($\tilde{C}_l^{\rm obs} =  \tilde{C}_l^{\rm th} + E_l^{ij}$), the minimization gives (up to a very good approximation) $\epsilon_l=L^{-1/2}$, and 
\begin{equation}
\Delta \chi^2_{\rm eff} = \sum_l  \left[(2l+1) f_{\rm sky} \left(N + 0 - N\right)  + L^{-1}\right] = 1.
\end{equation}

\subsection{Accounting for an extra neutrino-related error\label{B5}}

Finally, we account for the correlated error modelling neutrino-related uncertainties by multiplying the theoretical power spectrum $P^{\rm th}(k,z)$ by a factor $(1+e_\nu \sigma_\nu(k,z))$, as in equation (\ref{eq:e_nu}), as well as adding $e_\nu^2$ to $\Delta \chi^2_{\rm eff}$. The nuisance parameter $e_\nu$ is then marginalized over. Note that the factor $(1+e_\nu \sigma_\nu(k,z))$ should not multiply the observed/fiducial spectrum, as long as we assume a fiducial value of $e_\nu$ equal to zero.

\bibliographystyle{utcaps}
\bibliography{euclid}


\end{document}